\documentclass[twocolumn,aps,prd,preprintnumbers,nofootinbib,showpacs,floatfix,showkeys]{revtex4-1} 

\usepackage[usenames,dvipsnames]{xcolor} 
\usepackage{amsmath,amssymb,amsxtra} 
\usepackage{epsfig,subfigure} 
\usepackage[colorlinks=false]{hyperref}

\definecolor{fu-grey}{rgb}{0.36, 0.35, 0.30} 
\definecolor{fvb}{RGB}{204, 0, 0} 
\definecolor{jn}{RGB}{10, 10, 200} 
\definecolor{pgf}{RGB}{10, 150, 10} 

\newcommand*{\mathcolor}{}
\def\mathcolor#1#{\mathcoloraux{#1}}
\newcommand*{\mathcoloraux}[3]{%
  \protect\leavevmode
  \begingroup
    \color#1{#2}#3%
  \endgroup
}
\newcommand{\textins}[2][fu-grey]{
    \ifmmode\mathcolor{#1}{#2}
    \else\textcolor{#1}{#2}\@\,
    \fi
}

\makeatletter
\newcommand*{\cf}{\textit{cf}.\@\,}
\newcommand*{\eg}{\textit{e}.\textit{g}.\@\,}
\newcommand*{\ie}{\textit{i}.\textit{e}.\@\,}
\newcommand{\etc}{\@ifnextchar{.}{\&\textrm{c}}{\&\textrm{c}.\@}}
\newcommand{\etal}{\@ifnextchar{.}{\&\textrm{al}}{\&\textrm{al}.\@}}
\makeatother

\newcommand{\qsa}{quasi-static approximation}

\newcommand{\vect}[1]{\vec{#1}} 
\newcommand{\Exp}[1]{\mathrm{exp} \left( {#1} \right)} 
\newcommand{\abs}[1]{\lvert#1\rvert}

\makeatletter
\newcommand*\widet[1]{\mathpalette\wthelper{#1}}
\newcommand*\wthelper[2]{%
        \hbox{\dimen@\accentfontxheight#1%
                \accentfontxheight#1 1.2\dimen@
                $\m@th#1\widetilde{#2}$%
                \accentfontxheight#1\dimen@
        }%
}
\newcommand*\accentfontxheight[1]{%
        \fontdimen5\ifx#1\displaystyle\textfont
        \else\ifx#1\textstyle\textfont
        \else\ifx#1\scriptstyle\scriptfont
        \else\scriptscriptfont
        \fi\fi\fi3
}
\makeatother
\newcommand{\xtilde}[2][]{\ensuremath\widet{#2}}
\newcommand{\dtilde}[1]{\dot{\xtilde{#1}}}
\newcommand{\ddtilde}[1]{\ddot{\xtilde{#1}}}

\newcommand{\nn}{\nonumber}
\newcommand{\hub}{\ensuremath\mathcal{H}}
\newcommand{\comment}[1]{}

\graphicspath{{./UpdatedPlots/}}
\newcommand{\figref}[1]{Figure~\ref{#1}}
\newcommand{\eqnref}[1]{Equation~\ref{#1}}
\newcommand{\secref}[1]{Section~\ref{#1}}

\begin{document}
\title{Relativistic scalar fields and the quasi-static approximation \\  in theories of modified gravity}
\date{\today}
\author{Johannes Noller}
\email{noller@physics.ox.ac.uk}
\author{~Francesca von~Braun-Bates} 
\author{Pedro G. Ferreira}
\affiliation{Astrophysics, University of Oxford, DWB, Keble Road, Oxford, OX1 3RH, UK}

\begin{abstract}

Relativistic scalar fields are ubiquitous in modified theories of gravity. An important tool in understanding their impact on structure formation, especially in the context of N-body simulations, is the quasi-static approximation in which the time evolution of perturbations in the scalar fields is discarded.
%
%
We show that this approximation must be used with some care by studying linearly perturbed scalar field cosmologies and quantifying the errors that arise from taking the quasi-static limit. We focus on  $f(R)$ and chameleon models and link the accuracy of the quasi-static approximation to the fast/slow-roll behaviour of the background and its proximity to $\Lambda$CDM. 
Investigating a large range of scales, from super- to sub-horizon, we find that slow-rolling ($\Lambda$CDM-like) backgrounds generically result in good quasi-static behaviour, even on (super-)horizon scales.
We also discuss how the approximation might affect studying the non-linear growth of structure in numerical N-body  simulations. 
\end{abstract}

\maketitle

\section{Introduction}
\label{intro}
While relativistic scalar fields are hard-wired into our current theories of the very early universe, they are also at the heart of our modern understanding of the evolution of the universe at late times \cite{Copeland:2006wr}.  They are often invoked as the source of dark energy as well as being instrumental in attempts at modifying general relativity \cite{Clifton:2011jh}. As such, their presence should be felt and have a significant impact on the formation of structure.

The role that relativistic scalar fields play in linear cosmological perturbations of homogeneous universes is well-developed and understood. From coherent perturbations as one finds in a wide range of Quintessence \cite{FerreiraJoyce,Hu1998,PerrottaBaccigalupi1999} models to incoherent perturbations as emerge in axion and axion-like theories \cite{TurnerWilczekZee83,AmendolaBarbieri2006,MarshFerreira2010,MarshGrinHlozekFerreira2013} , it is now possible to calculate cosmological observables in the linear regime with almost arbitrarily high precision. Furthermore, a range of phenomenological approaches exist which can be applied to understand the effects of the scalar field in different ways. 

The same cannot be said on small scales where non-linear effects come into play. There, the method of choice is to use N-body simulations to study how non-linear evolution will lead to the formation of galaxies, clusters and, more generally, the cosmic web that is such a rich source of dynamical information.  N-body simulations are inherently {\em non-relativistic} --- for they simulate a system which interacts under Newtonian gravity --- and as such should not, in principle, capture the essential relativistic nature of the scalar field. While there have been attempts at inserting scalar fields into N-body simulations, in general they have been at the expense of taking the equivalent Newtonian limit of the scalar field equation of motion \cite{1303.0007,0807.2449}. Broadly speaking this means converting a dynamical, sourced, Klein-Gordon equation into a Poisson-like equation: the {\em quasi-static} approximation (we will explain this approximation more thoroughly later).

The usefulness of the \qsa{} and consequently its wide-spread use (consider for example the N-body codes \cite{Li:2011vk,Puchwein:2013lza,Llinares:2013jza}) stem from the fact that evaluating the full unapproximated evolution equations in N-body simulations is a task which is often computationally expensive. An illustrative example are chameleon scenarios where evaluating the full evolution equations quickly leads to computations requiring $\sim {\cal O}(10^7)$ more time steps than their quasi-static counterparts or more\cite{1303.0007}. 
In $f(R)$ models N-body simulations implementing the quasi-static approximation have been carried out e.g. by \cite{0807.2449,Koyama:2009me,Schmidt:2010jr,Ferraro:2010gh,Lam:2012by,Jennings:2012pt}\footnote{Interestingly the recent work of \cite{Adamek:2013wja} outlines a different simulation strategy not explicitly relying on quasi-static behaviour and which should be applicable to models with relativistic scalars in the future.}, see especially \cite{0807.2449}  for a numerical check of the \qsa{} in this context. For related chameleon models \cite{KhouryWeltman} also see \cite{Lombriser:2013wta,Brax:2013mua}. Non-linear structure formation in braneworld-inspired DGP models \cite{Dvali:2000hr} has been probed by  \cite{Laszlo:2007td,Khoury:2009tk,Schmidt:2009sg,Chan:2009ew,Schmidt:2009yj,Koyama:2009me,Schmidt:2010jr}, where 
\cite{Schmidt:2009sg} concludes that the \qsa{} is a self-consistent approach on sub-horizon scales in this setup. Linear \cite{Barreira:2012kk} and non-linear \cite{Barreira:2013xea, Barreira:2013eea, Li:2013tda} structure formation for galileon models \cite{Nicolis:2008in} have also been probed. Interestingly there the \qsa{} may break down particularly in low density regimes. In the linearised regime, however, it generically performs well on sub-horizon scales \cite{Barreira:2012kk}. 
    
While the \qsa{} therefore appears to do reasonably well in a number of model-specific contexts and there are very good arguments for its general 'reasonableness' in known observationally viable modified gravity models \cite{Silvestri:2013ne},
there are also known cases where it explicitly breaks down even on sub-horizon scales \cite{0802.2999,1302.1774}. Note, however, that it is not quite clear whether any of those known non-quasi-static scenarios have clear observational signatures in allowed regions of parameter space\footnote{We thank Claudio Llinares and Alessandra Silvestri for bringing this point to our attention.}
In this context also especially note the work of \cite{0802.2999}, which links the applicability of the \qsa{} on sub-horizon scales in $f(R)$ models to the proximity of the background evolution to $\Lambda$CDM\footnote{More precisely, the condition is $\abs{\partial_R f(R)} \ll 1$ at all times.  The present-day value of $\partial_R f(R)$ is abbreviated $f_{R0}$. In particular this means that large classes of observationally viable $f(R)$ theories, \ie{} those falling within the constraint $\abs{f_{R0}} \leq 10^{-6}$ imposed by a combination of solar-system and galaxy-halo tests \cite{0705.1158}, should satisfy a number of constraints \cite{0802.2999,1210.6880} guaranteeing good quasi-static behaviour.} and also \cite{1210.6880} who also probe linear growth in $f(R)$ theories in the quasi-static approximation.
The \qsa{} is also extended to Jordan-Brans-Dicke theories in \cite{Cembranos:2013bka} and to $f(R,T)$ models in \cite{1302.1866}\footnote{The scalars $R$ and $T$ are the Ricci scalar and the trace of the stress-energy tensor respectively}, where the inclusion of an $f(T)$ term causes scale-dependent behaviour of the density oscillations (in both the unapproximated equations and the quasi-static limit), resulting in inaccurate quasi-static behaviour. In general, and particularly for non-linear structure formation, however, the de facto necessity of the approximation in numerical modelling makes it inherently difficult to precisely determine its range of validity.

%
%
%
Our approach, in this paper, is to explore the validity of the quasi-static approximation on both large and small scales by using the apparatus of linear perturbation theory. In order to do so, we perform a detailed comparison between quasi-static and full, not approximated evolutions. The models which we consider are representative $f(R)$ and chameleon models of modified gravity, which alternatively may be interpreted as $f(R)$ models without and with screening. Doing so we aim to extend previous work by analytically and quantitatively understanding on which scales and subject to what conditions exactly the \qsa{} is a valid approximation for both $f(R)$ and chameleon models.
We explore and quantify these models in enough detail that we can use our results as a guide on how to tackle and better understand the evolution of non-linear perturbations in N-body simulations in the future. In doing so we identify the regimes where the quasi-static approximation can and cannot be trusted.  

This paper is structured as follows. In \secref{CP} we lay out the pared-down formalism of cosmological perturbations which we will use throughout the paper and in \secref{QS} we use it to understand the Newtonian limit, the quasi-static approximation and the miracle of N-body simulations with non-relativistic matter, which does not extend to relativistic scalar fields. 
In \secref{FR} we then apply the quasi-static approximation to $f(R)$ models with an exponential potential and compare it to the full evolution of perturbations without the \qsa{}. Providing explicit examples, in \secref{EG} we map out the regime of validity of the quasi-static approximation and how it relates to the fast- and slow-rolling nature of the background scalar degree of freedom as well as its proximity to $\Lambda$CDM-like behaviour. This analysis is extended to specific $f(R)$ models with screening, namely chameleons, in \secref{SCR}.  Finally, in \secref{D} we discuss our findings and conclude.

\section{Cosmological perturbations}
\label{CP}
Throughout this paper we will use linear, cosmological perturbation theory to gain insight into structure formation in modified gravity.
%
%
 To do so, we need to perturb the metric and the energy content of the universe around a homogeneous and isotropic background.
Depending on one\rq{}s educational background (see \cite{BertschingerMa1994} for a thorough discussion), one tends to pick one of two gauges: synchronous or conformal Newtonian.  In the {\it synchronous} gauge one chooses a foliation of space-time such that surfaces of equal time correspond to those of equal density --- consequently the coordinates are those of a freely falling observer --- and the metric can be written
\begin{eqnarray}
ds^2=a^2(\tau)[-d\tau^2+(\gamma_{ij}+h_{ij})dx^idx^j] \nonumber \label{syncheq}
\end{eqnarray}
where $\tau$ is conformal time, $a(\tau)$ is the scale factor, $\gamma_{ij}$ is the conformal 3-space metric of constant Gaussian curvature and $h_{ij}$ its perturbation (from the Fourier-space parametrisation of  the scalar modes we have $h_{ij}=h\delta_{ij}/3+(h+6\eta)({\hat k}_i {\hat k}_j-\delta_{ij}/3)$ {where} ${\hat k}_i$ is the unit vector in the direction of the wave vector ${\vec k}$). 
Alternatively in the {\it conformal Newtonian} gauge, the metric is diagonal such that
\begin{eqnarray}
ds^2=a^2(\tau)[-(1+2\Psi)d\tau^2+(1-2\Phi)\gamma_{ij}dx^idx^j] \nonumber \label{cneq}
\end{eqnarray}
where $\Phi$ and $\Psi$ map directly on to the Newtonian potential field in the non-relativistic limit.
In this paper we will primarily work with the synchronous gauge, although we will resort to the conformal gauge to make a few key points.

The content of the universe must also be suitably perturbed so that key tensors retain a gauge-invariant structure.  For example, the stress energy of a perfect fluid has for its $(0,\mu)$ components:
\begin{eqnarray}
T^{0}_{\phantom{0}0}&=&-\rho(1+\delta) \nonumber \\
ik^jT^{0}_{\phantom{0}j}&=&(\rho+P)\theta \nonumber
\end{eqnarray}
where $\rho$ and $P$ are the background energy density and pressure, $\delta$ and $\theta$ are the density contrast and the momentum (the divergence of the 3-velocity perturbation) and we have transformed to Fourier space assuming the convention of \cite{BertschingerMa95}. While the structure of the perturbed energy momentum tensor is identical in both gauges, the perturbation variables $\delta$ and $\theta$ behave differently in both gauges. So for example, in synchronous gauge, the evolution of $\delta$ and $\theta$ for a pressure-less fluid is given by
\begin{eqnarray}
{\dot \delta}&=&-\theta-\frac{\dot h}{2} \nonumber \\
{\dot \theta}&=&-{\hub}\theta \nonumber
\end{eqnarray}
while in conformal Newtonian gauge we have
\begin{eqnarray}
{\dot \delta}&=&-\theta-3{\dot \Phi} \nonumber \\
{\dot \theta}&=&-{\hub}\theta+k^2\Psi \nonumber
\end{eqnarray}
where we have used the conformal Hubble factor, $\hub =\frac{\dot a}{a}$ and ${\dot a}=\frac{da}{d\tau}$.

To determine the perturbed metric (and close the system of equations), one needs to consider the perturbed Einstein field equations, $\delta G^{\alpha}_{\phantom{\alpha}\beta}=8\pi G \delta T^{\alpha}_{\phantom{\alpha}\beta}$ where $\delta G^{\alpha}_{\phantom{\alpha}\beta}$ and $\delta T^{\alpha}_{\phantom{\alpha}\beta}$ are the perturbed Einstein and energy-momentum tensor. In the conformal Newtonian gauge, we can combine the ($0$,$\beta$) components to construct the relativistic Newton-Poisson equation:
\begin{eqnarray}
-k^2\Phi=4\pi G a^2\left(\delta T^{0}_{\phantom{0}0}-3\frac{\hub}{k^2}ik^i \delta T^{0}_{\phantom{0}i} \right) \label{GNP}
\end{eqnarray}
In the synchronous gauge we have that
the metric is found by solving:
\begin{eqnarray}
k^2\eta-\frac{1}{2}{\hub}{\dot h}=-4\pi Ga^2\delta T^{0}_{\phantom{0}0} \nonumber \\
{\ddot h}+2{\hub}{\dot h}-2k^2\eta=-8\pi G a^2 \delta T^{i}_{\phantom{i}i} \nonumber
\end{eqnarray}
Specialising to the case of a shear-free fluid, we have 
\begin{eqnarray}
\delta T^{i}_{\phantom{i}j}=\delta P\delta^{i}_{\phantom{i}j} \nonumber
\end{eqnarray}

Finally, it makes sense to reduce the contents of the universe to a scalar field and dust, where the dust mimics dark matter and the scalar field is the 'modified gravity/dark energy degree of freedom'\footnote{Note that in effect this means we will be considering accelerating models that start in a matter-dominated regime and transition into one dominated by the scalar. We do not include the effect of radiation throughout this paper.}.  We now consider the evolution and effect of a scalar field, the archetypal relativistic source in modern cosmology. We will consider models with more complicated matter-scalar field couplings later on, but for the moment it is instructive to focus on a simple example of a Quintessence-like model where matter and the scalar are minimally coupled to gravity without any direct coupling to one another \cite{FerreiraJoyce}. Typically a scalar field $\varphi$ obeys a relativistic Klein-Gordon equation
\begin{eqnarray}
\nabla^\mu\nabla_\mu \varphi=-\frac{dV}{d\varphi} \nonumber
\end{eqnarray}
The scalar field can be divided into homogeneous and inhomogeneous components $\varphi=\phi+\chi$ which satisfy 
\begin{eqnarray}
{\ddot \phi}+2{\hub}{\dot \phi}+a^2V'=0 \label{phievol}
\end{eqnarray}
where $V'=dV/d\phi$ and
\begin{eqnarray}
{\ddot \chi}+2{\hub}{\dot \chi}+k^2\chi+a^2V''(\phi)\chi={\mathcal{S}} \label{chievol}
\end{eqnarray}
where ${\mathcal{S}}=-\frac{1}{2}{\dot \phi}{\dot h}$ in the synchronous gauge and ${\mathcal{S}}= 4{\dot \phi}{\dot\Phi}-2a^2V'\Phi$ in conformal Newtonian gauge.
The perturbed stress energy components for a scalar field are now
\begin{eqnarray}
\delta T^{0}_{\phantom{0}0}&=&-a^{-2}{\dot \phi}{\dot \chi}-V'(\phi)\chi \nonumber  \\
ik^i\delta T^{0}_{\phantom{0}i}&=& a^{-2}{\dot \phi}k^2\chi \nonumber \\
\delta T^{i}_{\phantom{i}i}&=&  a^{-2}{\dot \phi}{\dot \chi}-V'(\phi)\chi \nonumber 
\end{eqnarray}
We can combine these equations to obtain a coupled set of $2^{\rm nd}$ order ordinary differential equations in Fourier space:
\begin{eqnarray}
\ddot \delta + {\hub} \dot \delta - \frac{3}{2}{\hub}^2 \Omega_m \delta - 2\dot\phi \dot\chi + a^2 V' \chi &=& 0 \nonumber \\
\ddot\chi + 2 {\hub} \dot\chi + k^2 \chi + a^2 m^2_\phi \chi - \dot\phi \dot\delta &=& 0 \label{Qeom}
\end{eqnarray}
where $m^2_\phi=d^2V/d\phi^2$. In what follows, we will make use of these equations in exploring the evolution of cosmological perturbations in the linear regime and also re-encounter them in the context of $f(R)$.

\section{The quasi-static approximation and relativistic scalar fields}
\label{QS}

In this section we discuss a few aspects of cosmological perturbation theory and how we can use it as a guide to understanding N-body simulations of structure formation and the \qsa{}.  Let us first focus on \eqnref{GNP} and consider the case of a generic, perfect fluid with equation of state $w\equiv P/\rho$.  The Poisson equation in Fourier space is now
\begin{eqnarray}
-k^2\Phi=4\pi G a^2 \rho\delta_{gi} \label{RNP}
\end{eqnarray}
where we have defined the gauge-invariant density contrast
\begin{eqnarray}
\delta_{gi}\equiv \delta+\frac{3(1+w){\hub}}{k^2}\theta \nonumber
\end{eqnarray}

This is an interesting expression for a number of reasons. For a start, it differs from the non-relativistic Newtonian equation although in the limit where ${\mathcal{H}}/k \rightarrow 0$, namely on sub-horizon scales, they agree. Hence, in the Newtonian gauge, one expects relativistic corrections once one looks at sufficiently large scales. But more relevant is the fact that $\delta_{gi}$ is a gauge-invariant quantity and the relativistic Newton-Poisson equation we present above is gauge-invariant. The standard gauge-invariant Newtonian potentials map (by construction) directly on the conformal Newtonian potentials and, if accordingly we calculate $\delta$ and $\theta$ in any gauge, we can combine them to find $\delta_{gi}$. 

It turns out that this form of relativistic Newton-Poisson equation is at the heart of why N-body simulations can accurately calculate the evolution of the Universe from super-horizon down to sub-horizon scales, even though they, in principle, use the non-relativistic Newton-Poisson equation \cite{Bertschinger:1995er}. To understand why this is so, let us briefly sketch the algorithm for an N-body code. The idea is that one follows the motion of a set of N-particles (labelled by $a=1,\cdots N$) with positions $\vect{x}_a$. These particles obey the non-relativistic geodesic equation
\begin{eqnarray}
\frac{d^2 \vect{x}_a}{d\tau^2}+{\hub}\frac{d\vect{x}_a}{d\tau}=-\nabla\Phi(\vect{x}_a) \nonumber
\end{eqnarray}
while $\Phi$ is calculated (using a variety of integral techniques) from the non-relativistic equation:
\begin{eqnarray}
-k^2\Phi=4\pi G a^2 \rho\delta \label{simplePoisson}
\end{eqnarray}
Given that, na\"{i}vely, $\delta_{gi}\ne\delta$, one would expect that this equation is not applicable on scales of order the horizon or greater. Yet, it turns out that the $\delta$ as calculated in N-body simulations is in the frame of freely falling observers and hence in the synchronous gauge. If we now take the evolution equation for $\theta$ in that gauge, we see that it is solved by $\theta\propto a^{-1}$. Any initial perturbation in $\theta$ set up at early times will have completely died away and cannot be sourced at the linear level. This means that, in the synchronous gauge, $\delta_{gi}=\delta$. Given that $\Phi$ maps directly onto the gauge-invariant Newtonian potential, for a pressure-less fluid, 
\eqnref{simplePoisson} is therefore applicable on all scales. 

There are two major caveats in our explanation of why conventional N-body algorithms are applicable on cosmological scales (see also \cite{ChisariZaldarriaga} for the importance of getting the initial value constraint correct). For a start, we have used linear theory while the whole point of N-body simulations is to understand non-linear gravitational collapse; yet we are trying to understand gravitational collapse on the scale of the horizon and there we expect the evolution of gravitational collapse to be accurately described in the linear regime. But more importantly, we have focused on the case of pressure-less matter which fairly represents the dark matter that one is simulating. If the fluid is not pressure-less and non-relativistic, this argument breaks down.  The evolution equations for $\delta$ and $\theta$ for a shear-free perfect fluid in synchronous gauge are now (cf. \cite{Christopherson:2012kw})
\begin{eqnarray}
{\dot \delta}&=&-(1+w)(\theta+\frac{\dot h}{2})-3{\hub}(c^2_s-w)\delta \nonumber \\
{\dot \theta}&=&-{\hub}(1-3w)\theta+\frac{c^2_s}{1+w}k^2\delta \nonumber
\end{eqnarray}
while in the conformal Newtonian gauge they are
\begin{eqnarray}
{\dot \delta}&=&-(1+w)(\theta-{\dot \Phi})-3{\hub}(c^2_s-w)\delta \nonumber \\
{\dot \theta}&=&-{\hub}(1-3w)\theta+\frac{c^2_s}{1+w}k^2\delta \nonumber
\end{eqnarray}
where $c^2_s$ is the sound speed of the fluid. Note that the Laplacian term will play an
important role if $c_sk/{\hub} \ge 1$. Furthermore if $w\ge 1/3$, $\theta$ will not decay, at least at the linear level, and may play a significant role in $\delta_{gi}$. Hence, the non-relativistic Newton-Poisson equations should not be applied on the scale of the horizon or greater.

\begin{figure*}[htbp] 
\begin{center}$
\begin{array}{cc}
\includegraphics[width=0.503\linewidth,trim=0cm 0cm 0cm 0.8cm,clip=true]{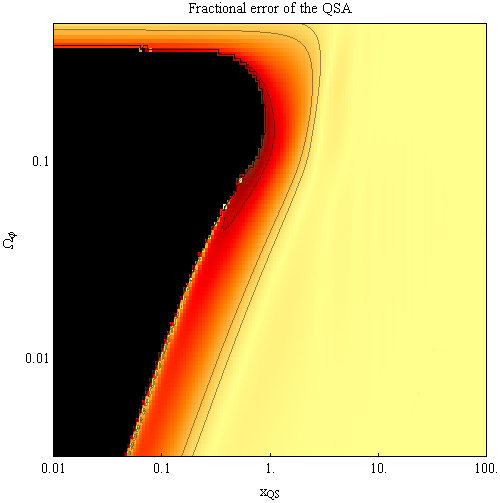} &
\includegraphics[width=0.497\linewidth,trim=0cm 0cm 0cm 0.6cm,clip=true]{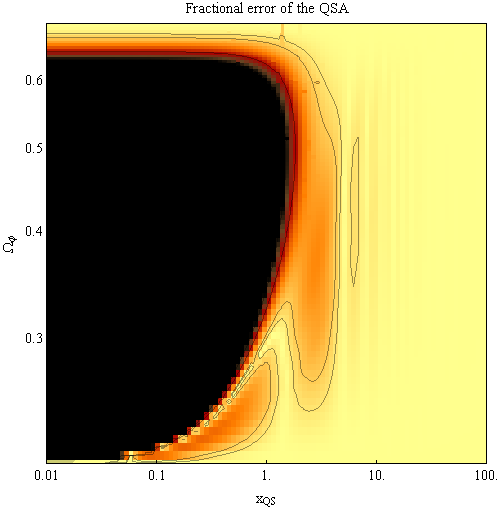} 
\end{array}$
\end{center}
\begin{center}$
\begin{array}{ccc}
\includegraphics[width=0.13\linewidth]{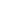} &
\includegraphics[width=0.7\linewidth]{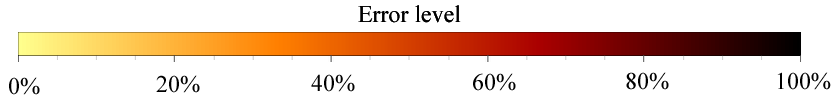} &
\includegraphics[width=0.05\linewidth]{whitespace.png}
\end{array}$
\end{center}
\caption{ Here we show the relative error $\delta_{QS}/\delta_{full} - 1$ resulting from the QSA in $f(R)$ for an accelerating, non-scaling regime ($\lambda = 1.5$ for this plot).
The x-axis denotes the value of $x_{QS} = k \tau_{QS}$, where $\tau_{QS}$ is the time when the \qsa{} is switched on.
The y-axis denotes the value of  $\Omega_\phi (\tau_{QS})$. The evolution is stopped and errors are computed once we reach $\Omega_{\phi(final)} = 0.7$. Note that the maximum value of the relative error increases with $\Omega_{\phi(final)}$, i.e. had we chosen $\Omega_{\phi(final)} > 0.7$ the errors shown would increase. Contours denote $5,10$ and $50 \%$ errors from right to left and the black region corresponds to $> 100 \%$ error. \textsc{Left:} A fast-roll $f(R)$ scenario, where the initial value of $\Omega_\phi$ is small and the field quickly starts evolving. \textsc{Right:} A slow-roll $f(R)$ setup with a large initial $\Omega_\phi$, where the field remains frozen-in ('slow-rolling') for a significant amount of time, cf. figure \ref{OmegaCompare}. The oscillatory features mildly visible on (sub-)horizon scales are a consequence of the oscillating behaviour of $\chi$ on those scales, cf. Figure \ref{deltaplots}.
\label{fRContour}}
\end{figure*}

A notable example is that of the relativistic scalar field introduced in the previous section. The relativistic Newton-Poisson equation is now
\begin{eqnarray}
-k^2\Phi=4\pi G a^2 \rho\delta_{gi}+4\pi G\left[{\dot \phi}{\dot \chi}+V'\chi+3{\hub}{\dot \phi}\chi\right] \label{ScaNP}
\end{eqnarray}
where the last term is the relativistic correction. In fact, we can see from Equations \eqref{phievol},\eqref{chievol} and \eqref{ScaNP}  that this system is fundamentally relativistic (Quintessence-like models have $c_s = 1$).  It seems, therefore that to accurately simulate a universe with the usual cosmological fluids and a relativistic scalar field it is necessary to evolve the full relativistic set of equations. This means that for an N-body simulation, not only is it necessary to solve the Newton-Poisson equation and the non-relativistic geodesic equation but also the second order evolution equations for $\phi$ and $\chi$. This is especially true if one wants to follow the evolution of modes that start off outside the cosmological horizon.  

There is a growing interest in simulating N-body systems in the presence of relativistic scalar fields and, as discussed in the introduction, the strategy in the overwhelming majority of cases has been to use the \qsa{} (henceforth \textit{QSA}) when evolving perturbations, where one assumes that\footnote{Our notation follows that of \cite{1109.2082} here.}
\begin{eqnarray}
{\abs{\nabla^2 X} \gg {\mathcal{H}}^2 \abs{X} \quad \text{and} \quad \abs{\dot{X}} \le {\mathcal{H}} \abs{X}}, \label{QSA1}
\end{eqnarray}
where \eg{} $X = \chi,{\dot \chi},h,\eta,..$ in synchronous gauge.
This approximation should be valid on sufficiently small (\ie{} sub-horizon) scales: indeed, it is remarkably efficient for evolving cosmological systems without actually having to follow the detailed evolution of the scalar field.  It is the purpose of this paper to explore how accurate this approximation actually is for a range of models which include a relativistic scalar field. Let us briefly summarise what exactly the QSA entails.
In essence it contains two separate assumptions as discussed \eg{} in \cite{Silvestri:2013ne}: 

(a) The relative suppression of time derivatives of metric/field perturbations compared with their spatial derivatives. 
\begin{equation}
\abs{\dot{X}} \le {\hub} \abs{X}
\end{equation}
Here we will solely be concerned with testing the validity of the quasi-static approximation as applied to scalar field fluctuations, so $X = \chi,{\dot \chi}$. 
In principle scalar field (as well as matter and metric) perturbations can follow an evolution with non-negligible time-derivatives, \eg{} by displaying highly oscillatory behaviour. However, typically these are heavily constrained. For example, in the case of $f(R)$ gravity $\leftrightarrow$ chameleon models it has been argued that the relative suppression of such derivatives, effectively a slow-roll condition for  $\dot{\phi}$, is required by solar system constraints (in order to have a successful screening of fifth forces) \cite{KhouryWeltman,Brax:2012gr,0705.1158}. One should keep in mind, however, that this is a model-dependent statement - see \eg{} \cite{1302.1774} for a symmetron model with collapsing domain walls; a feature absent if a \lq{}static\rq{} simulation is employed.

(b) A sub-horizon approximation $k^2 \gg \hub^2$ or, when written in the same formalism as above 
\begin{equation}
\abs{\nabla^2 X} \gg {\mathcal{H}}^2 \abs{X},
\end{equation}
where as above we will be concerned with the case when $X = \chi,{\dot \chi}$. This assumption is typically required, since ignoring time-derivatives amounts to neglecting any slow-varying changes to $\chi$ as well, which is only justified on sub-horizon scales, where $\chi$ has decayed away sufficiently, so that its evolution is no longer important.\footnote{The oscillatory features visible on (sub-)horizon scales in the contour plots \ref{fRContour} and \ref{chamcontplots} are a result of the intermediate phase where $\chi$ is displaying an oscillatory decay, but is still relevant. As a result these features vanish as $x_{QS}$ becomes large, i.e. as the field $\chi$ decays away.}
Also note that, in $\Lambda$CDM-like models, the evolution time scale for perturbations is set by the Hubble rate and consequently assumption (b) there entails (a).  

Having characterised the \qsa{} and how it is used in N-body simulations, we now proceed to explore a few representative models. In doing so, we identify the key qualitative features which make the \qsa{} a useful and and accurate tool.

\section{$f(R)$ gravity}
\label{FR}
In this section and the next we will compare the exact evolution of linearised perturbations in different types of $f(R)$ models with its quasi-static and hence approximate counterpart. The aim is to assess in what regimes the \qsa{} is a well-behaved approximation and in particular whether its naive range of validity (good on subhorizon scales, bad on superhorizon scales) can be extended.   

An $f(R)$ theory can be defined in the Jordan frame via the action 
\begin{equation}
S_J =\frac{1}{2}\int d^4 x\sqrt{-g}\, \left[R+f(R)\right] + \int d^4 x\sqrt{-g}\, {\mathcal{L}}_{\rm m}[\Phi_i,g_{\mu\nu}] \ ,
\label{fRaction}
\end{equation}
where we have chosen units such that $8\pi G = 1$, the function $f(R)$ is a general function of the Ricci scalar, $R$, and $\Phi_i$ denotes all matter fields. Via a series of field redefinitions and a conformal transformation \cite{Chiba:2003ir,Magnano:1993bd,Barrow:1988xh,0611321} we can turn the Jordan frame action into an equivalent Einstein frame one 
\begin{eqnarray}
\nn S_E &=&\frac{1}{2}\int d^4 x\sqrt{-\xtilde{g}}\, \xtilde{R} \\
&+& \int d^4 x\sqrt{-\xtilde{g}}\, 
\left[-\frac{1}{2}\xtilde{g}^{\mu\nu}\xtilde{\nabla}_{\mu}\phi \xtilde{\nabla}_{\nu}\phi -V(\phi)\right] \nonumber
\\ &+& S_{\text{matter}}[\Phi_i, e^{-\beta \phi} \xtilde{g}_{\mu\nu}]
\label{einsteinaction}
\end{eqnarray}
where a tilde denotes Einstein frame quantities and we have performed a conformal transformation
\begin{equation}
\xtilde{g}_{\mu\nu} = e^{2 \omega} g_{\mu\nu},
\label{conformal}
\end{equation}
requiring
\begin{eqnarray}
e^{-2 \omega}(1+f_{R}) &=& 1, \\
\phi &\equiv& \frac{2\omega}{\beta},
\end{eqnarray} 
where $f = f(R)$ and a subscript $R$ denotes differentiation w.r.t. $R$.
For $f(R)$ theories we have $\beta = \sqrt{2/3}$. The fact that we have this conformal transformation is the essential ingredient behind the mapping between $f(R)$ and chameleon-screened theories  \cite{KhouryWeltman}- we will return to this point in section \ref{SCR}. 
Finally the potential $V(\phi)$ is determined entirely by the original Jordan frame action and is given by
\begin{equation}
V(\phi)=\frac{1}{2}\frac{R f_{R} - f}{(1+f_{R})^{2}}.
\label{einsteinpotential}
\end{equation}

At this point one may wonder whether any particular fiducial form suggests itself for the potential. For an arbitrary polynomial of positive powers of $R$ in four dimensions of the form $\sum_{n=1}^{k} a_n R^n$, such a potential will asymptotically approach an exponential potential as $\phi \to \infty$. This is the fiducial potential chosen by \cite{FerreiraJoyce,0611321} and will be the potential we work with throughout most of this paper too. 
However, one may wonder what the relevant potential looks like for other motivated potentials of interest, \eg{} the Hu \& Sawicki model \cite{0705.1158}, where we have
\begin{equation}  
f(R) =R  -m^2 \frac{\left(c_1 \left(\frac{R}{m^2}\right)^n\right)}{1 + c_2 \left( \frac{R}{m^2}\right)^n}, \label{HSpot}
\end{equation}
where $c_1,c_2, n$ are arbitrary constants. We will return to the Hu \& Sawicki model in the context of the chameleon section \ref{SCR}, where we will also find that an exponential potential qualitatively is a good proxy for this model in several regions of parameter space. But for the time being we will continue to work in as much generality as possible without specifying a concrete potential. 

\begin{figure*}[htbp] 
\begin{center}$
\begin{array}{cc}
\includegraphics[width=0.5\linewidth]{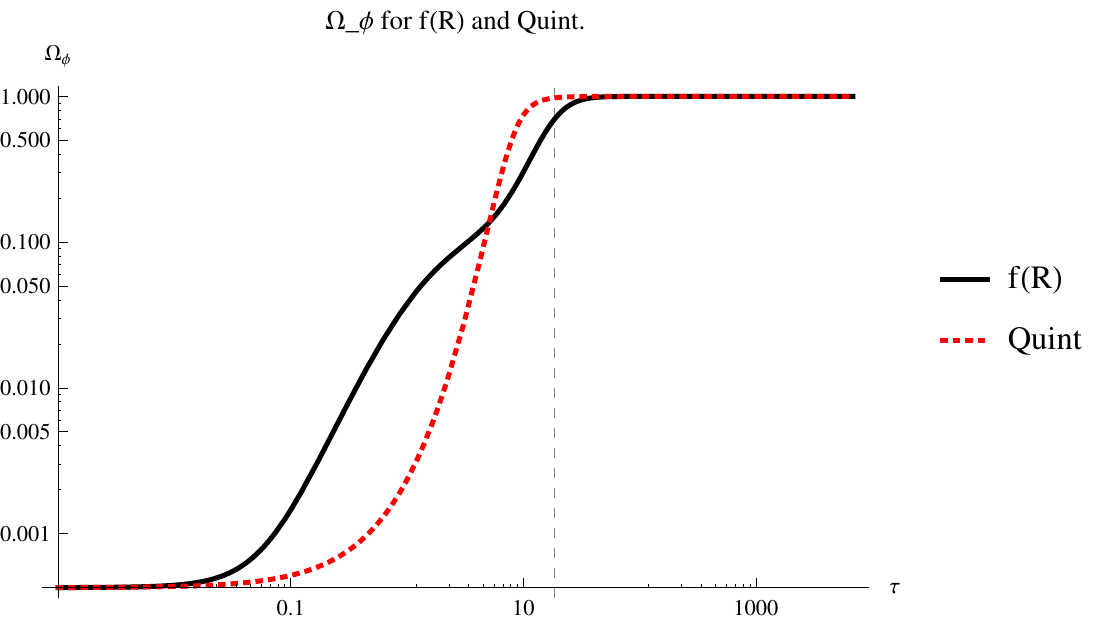} &
\includegraphics[width=0.5\linewidth]{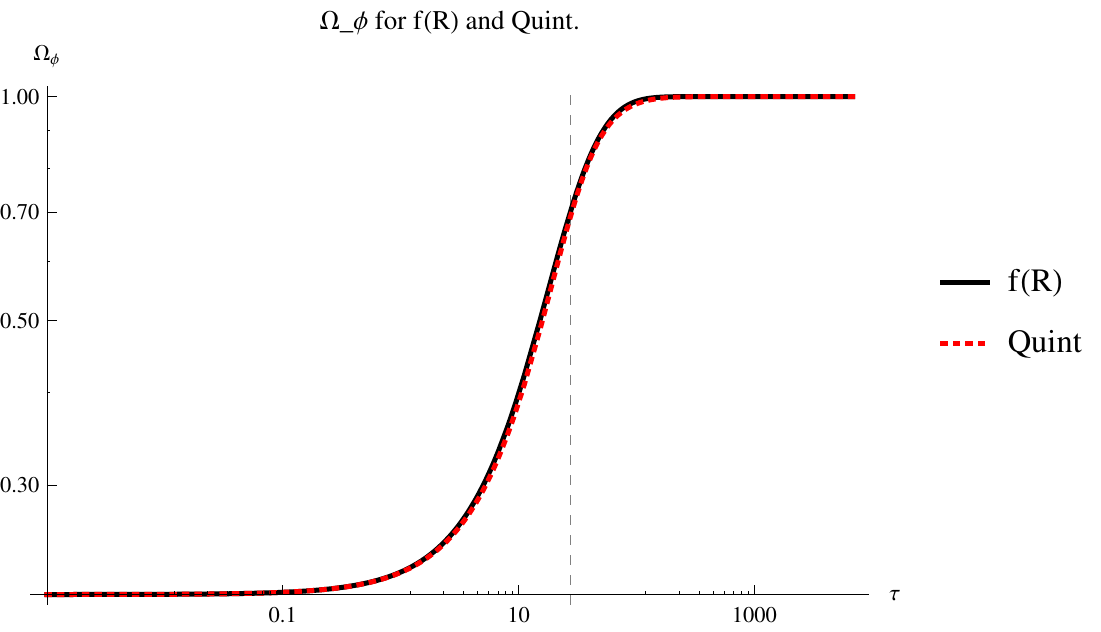}
\end{array}$
\end{center}
\caption{
The two different background evolutions in terms of $\Omega_\phi$ considered in this section. \textsc{Left}: Fast-roll $f(R)$ and corresponding Quint. evolutions starting with an initial $\Omega_{\phi,i} \sim 10^{-4}$ that quickly starts evolving in the $f(R)$ case. \textsc{Right}: Slow-roll $f(R)$ and corresponding Quint. evolutions starting with an initial $\Omega_{\phi,i} \sim 0.21$ that initially stays frozen in and only later starts evolving. $f(R)$ and Quint. evolutions are nearly indistinguishable in this case. Note that the vertical dashed lines indicates when $\Omega_\phi = 0.7$ in the $f(R)$ model considered and that the y-axis has a different range in the two plots.
\label{OmegaCompare}}
\end{figure*}
\begin{figure*}[htbp] 
\begin{center}$
\begin{array}{cc}
\includegraphics[width=0.5\linewidth]{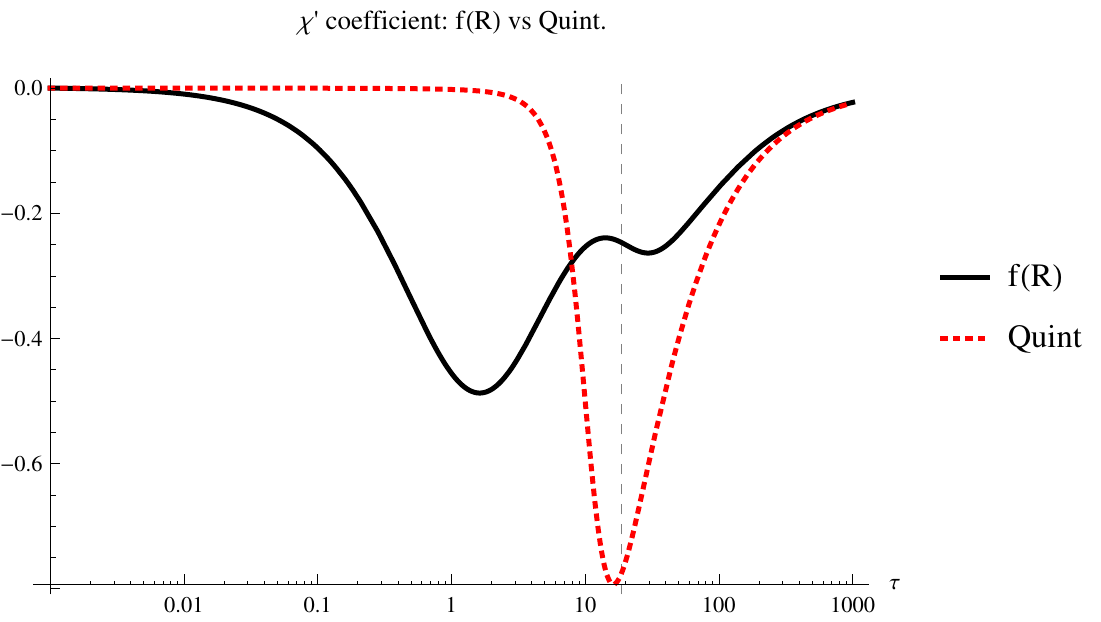} &
\includegraphics[width=0.5\linewidth]{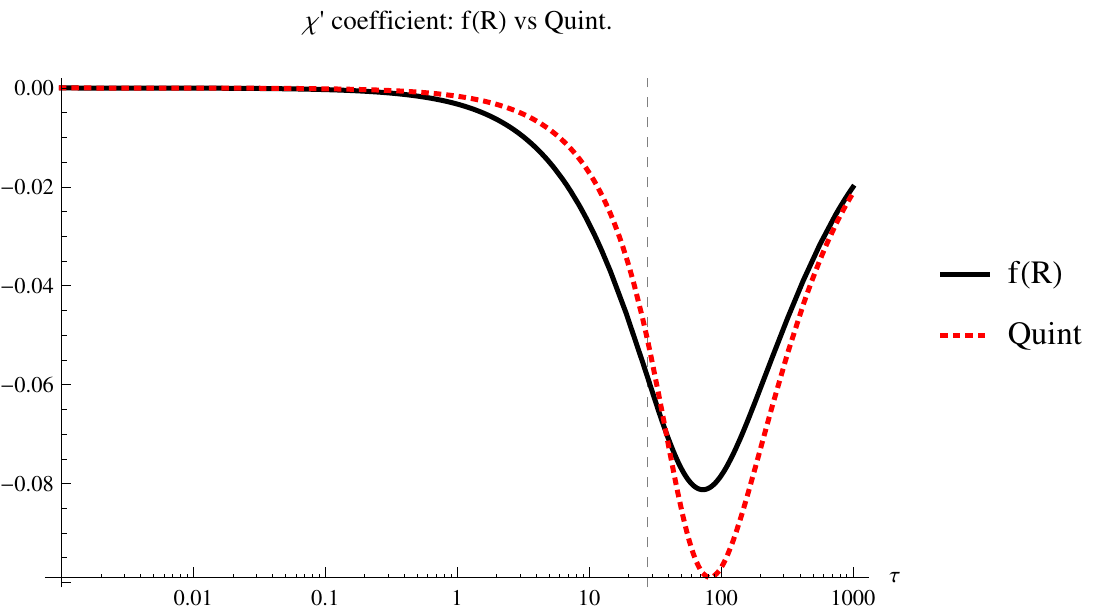} \\
\includegraphics[width=0.5\linewidth]{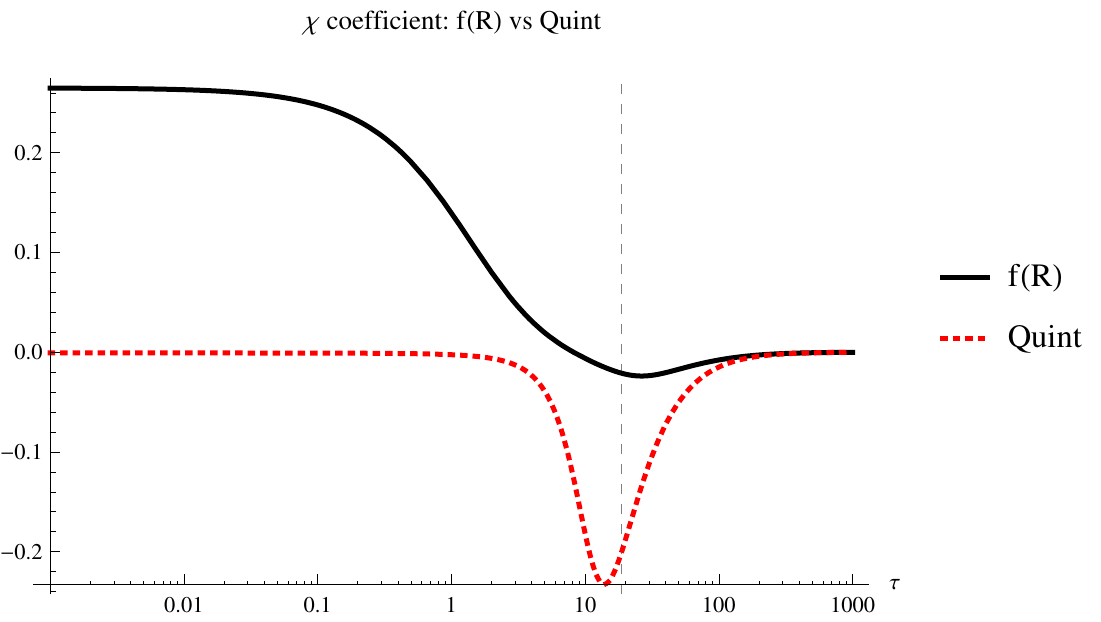} &
\includegraphics[width=0.5\linewidth]{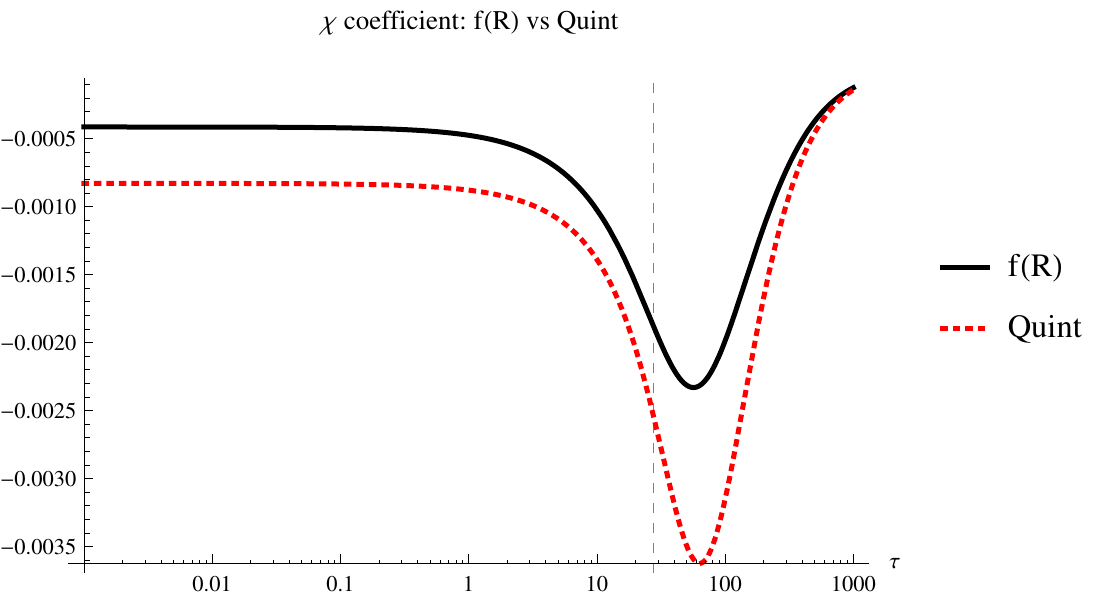}
\end{array}$
\end{center}
\caption{
Plots showing the evolution of the coefficients of $\chi$ (Equation \eqref{chico}) and $\dot{\chi}$ (Equation \eqref{chidotco}) for $f(R)$ and Quint. in the QSA evolution equations as discussed in Section \ref{FR}. Fast-roll cases are shown on the left, slow-roll on the right. Note that in the fast-roll case both coefficients are very small at early times when modes of interest are on (super-)horizon scales for Quint., while this is not the case for $f(R)$. In the slow-roll case coefficients are small for both $f(R)$ and Quint. leading to a suppression of the QSA error propagation. Once again the vertical dashed lines indicate when $\Omega_\phi = 0.7$ in the $f(R)$ models considered and hence the point at which errors are evaluated in the contour graphs \ref{fRContour}.
\label{chicoeff}}
\end{figure*}

The evolution of the background in an $f(R)$ model is governed by \cite{0611321}
\begin{eqnarray}
\xtilde{\hub}^{2} &=& \frac{1}{3}\left(\frac{\dot\phi^{2}}{2}+\xtilde{a}^{2} V(\phi)+\xtilde{a}^{2} \xtilde{\rho}_{m}\right) \nn
\\
\ddot\phi &+& 2 \xtilde{\hub}\dot \phi + \xtilde{a}^{2} V_{\phi}= \frac{1}{2}\beta \xtilde{a}^{2} \xtilde{\rho}_{m} \nn
\\
\xtilde{\rho}_{m} &\equiv &  \frac{ \xtilde{\rho}_{m}^{*0}} {\xtilde{a}^{3}} \exp\left(-\frac{\beta\phi}{2}\right). 
\label{fRb}
\end{eqnarray}
In synchronous gauge the perturbation equations are given by\footnote{The careful reader will have observed that there are two sign differences between equation \eqnref{chifr} and the analogous equation presented in \cite{0611321} - the version here corrects these typos.} 
\begin{align}
\ddtilde{\delta} + \xtilde{\hub} \dtilde{\delta}  -\frac{3}{2} \xtilde{\hub}^{2} \xtilde{\Omega}_{m}(\xtilde{\delta} -\frac{\beta\chi}{2}) - 2\dot\phi \dot\chi + \xtilde{a}^2 V_{\phi}\chi &= 0 \label{deltafr}
\\
\begin{aligned}[b]\ddot\chi +2 \xtilde{\hub} \dot\chi +k^{2}\chi
&+ \;\xtilde{a}^{2}V,_{\phi\phi}\chi - \dot\phi \dtilde{\delta} \\ 
&-  \;\frac{3 \beta}{2} \;\xtilde{\hub}^{2} \xtilde{\Omega}_{m}\;(\xtilde{\delta} -\frac{1}{2} \beta\chi) 
\end{aligned}
&= 0 \label{chifr}
\end{align} 
In the quasi-static approximation, the second perturbation equation can be used to solve for $\chi$, so that we now solve
\begin{eqnarray}
\nn \ddtilde{\delta} + \xtilde{\hub} \dtilde{\delta} -\frac{3}{2} \xtilde{\hub}^{2} \xtilde{\Omega}_{m}(\xtilde{\delta}-\frac{\beta\chi}{2})- 2 \dot\phi \dot\chi+ \xtilde{a}^2 V_{\phi}\chi 
&=& 0 \label{CDMeq2}
\\
\nn k^{2}\chi + \xtilde{a}^{2}V,_{\phi\phi}\chi - \dot\phi \dtilde{\delta} -  \frac{3\beta}{2} \xtilde{\hub}^{2} \xtilde{\Omega}_{m}(\xtilde{\delta} -\frac{1}{2} \beta\chi) &=& 0. \\ \label{QSFR}
&&
\end{eqnarray}
Application of the QSA eliminates $\dot{\chi}$, $\ddot{\chi}$ in \eqnref{chifr}, but not $\dot{\chi}$ in \eqnref{deltafr}, where there is no $k^2 \chi$ term relative to which $\dot{\chi}$ is suppressed. Note that, in the evolution equation for $\chi$, several terms survive the QSA. We have both a mass term as well as extra contributions dependent on $\dot\phi$ and $\delta$. 

In assessing the accuracy of the QSA in $f(R)$ models we will find it useful to compare them with analogous Quintessence-like solutions, \ie{} models with no non-minimal coupling to matter as present in the case of $f(R)$. This corresponds to setting $\beta=0$ in the action \eqref{einsteinaction}.  Consequently, the background evolution equations now are
\begin{eqnarray}
\xtilde{\hub}^{2} &=& \frac{1}{3}\left(\frac{\dot\phi^{2}}{2}+\xtilde{a}^{2} V(\phi)+\xtilde{a}^{2} \xtilde{\rho}_{m}\right) \nn
\\
\ddot\phi &+& 2 \xtilde{\hub}\dot \phi + \xtilde{a}^{2} V_{\phi}= 0 \nn
\\
\xtilde{\rho}_{m} &\equiv &  \frac{\xtilde{\rho}_{m}^{*0}}{\xtilde{a}^{3}}, \label{Quintb}
\end{eqnarray}
whereas perturbations are governed by
\begin{align}
\ddtilde{\delta} + \xtilde{\hub} \dtilde{\delta}  -\frac{3}{2} \xtilde{\hub}^{2} \xtilde{\Omega}_{m}\xtilde{\delta} - 2\dot\phi \dot\chi + \xtilde{a}^2 V_{\phi}\chi &= 0 \label{CDMeq}
\\
\begin{aligned}[b]\ddot\chi +2 \xtilde{\hub} \dot\chi +k^{2}\chi
+ \;\xtilde{a}^{2}V,_{\phi\phi}\chi - \dot\phi \dtilde{\delta}  \label{chiQeom}
\end{aligned}
&= 0
\end{align} 
and the quasi-static approximation reduces this to
\begin{align}
\nn \ddtilde{\delta} + \xtilde{\hub} \dtilde{\delta} -\frac{3}{2} \xtilde{\hub}^{2} \xtilde{\Omega}_{m} \xtilde{\delta}- 2 \dot\phi \dot\chi+ \xtilde{a}^2 V_{\phi}\chi &= 0 \label{Quinteq2}
\\
 k^{2}\chi + \xtilde{a}^{2}V,_{\phi\phi}\chi - \dot\phi \dtilde{\delta}  &= 0. 
\end{align}
Note how, by taking the limit $\beta \to 0$, Equations \eqref{CDMeq} and \eqref{chiQeom} have exactly reproduced the evolution equations for the simple Quintessence-like model in Equation \eqref{Qeom}.

\section{The fast and slow roll regime of $f(R)$}\label{EG}

It should already be obvious that there are some fundamental differences at the perturbative level between a Quintessence-like model (henceforth Quint.) and an $f(R)$ model as described in the previous section. To understand this difference, in particular in the context of the QSA, consider the solutions to the quasi-static evolution equations: 
\begin{align}
\chi_{QSA}^{Quint} &=  \frac{\dot\phi \dot\delta}{ k^2 + a^2 V_{,\phi\phi}  },\nonumber\\
 \chi_{QSA}^{f(R)} &=  \frac{\dot\phi \dot\delta + \frac{3}{2} \beta \hub^2 \Omega_m \delta} {k^2 + a^2 V_{,\phi\phi} + \frac{3}{4} \beta^2 \hub^2 \Omega_m  } \label{inacchi}
\end{align}
Our primary interest is the evolution of $\delta$ and errors introduced into this evolution by the QSA. These errors come from the fact that, in the QSA, we simplify the $\chi$ evolution equation and hence obtain an inaccurate solution for $\chi$ \eqref{inacchi} \footnote{This inaccuracy mainly appears on (super-)horizon scales. On sub-horizon scales the QSA does well by design (at least for the examples considered throughout this paper - for counterexamples see \cite{0802.2999,1302.1774}) and the corresponding $\chi$ solution is a faithful one.}.  This propagates to the evolution equation for $\delta$ via its direct dependence on $\chi$ as well as a dependence on $\dot\chi$ via the $\dot\phi \dot\chi$ term. How much of this error propagates determines how well the QSA does. However, already at this point it becomes clear that the slow- or fast-roll properties of the background (the size of $\dot \phi$) will be important for error propagation in the QSA. It will prove useful to consider two concrete $f(R)$ examples. We emphasize that we treat these examples as toy models in order to understand both qualitatively and quantitatively why and when the QSA does well - for the time being, we will therefore not be concerned with tuning all of the model parameters to match observational constraints, but focus on generic features of such models. We will comment on the observational viability of these toy models in the Section \ref{SCR}. For both example cases we will, as discussed in the previous section and following \cite{FerreiraJoyce,0611321}, pick an exponential potential of the form $V \sim \Exp{-\abs{\lambda}\phi}$, choosing $\lambda = 1.5$ so that we obtain a non-scaling, accelerating background solution in which the scalar field dominates at late times. The difference between the two cases will solely consist in the initial conditions imposed on the scalar field, leading to different background evolutions. 

\begin{figure*}[htbp] 
\begin{center}$
\begin{array}{cc}
\includegraphics[width=0.5\linewidth]{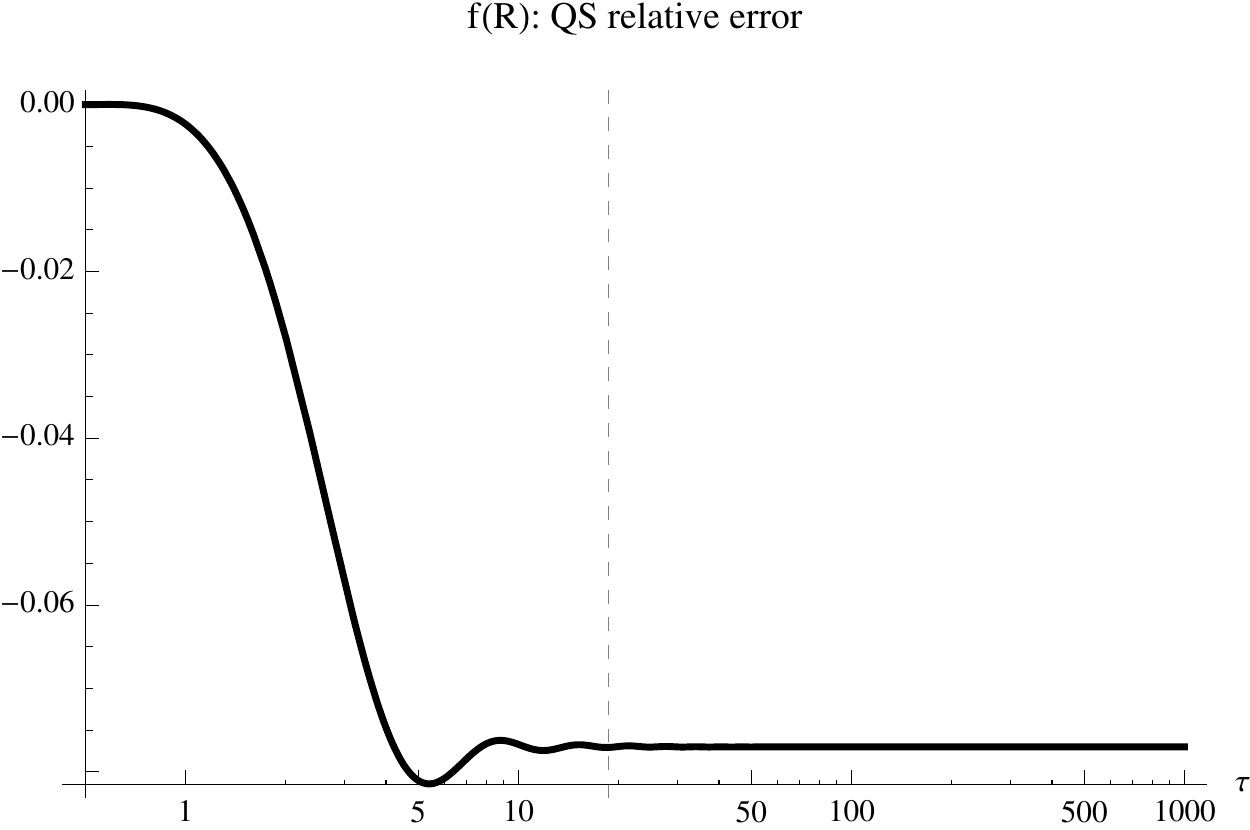}& 
\includegraphics[width=0.5\linewidth]{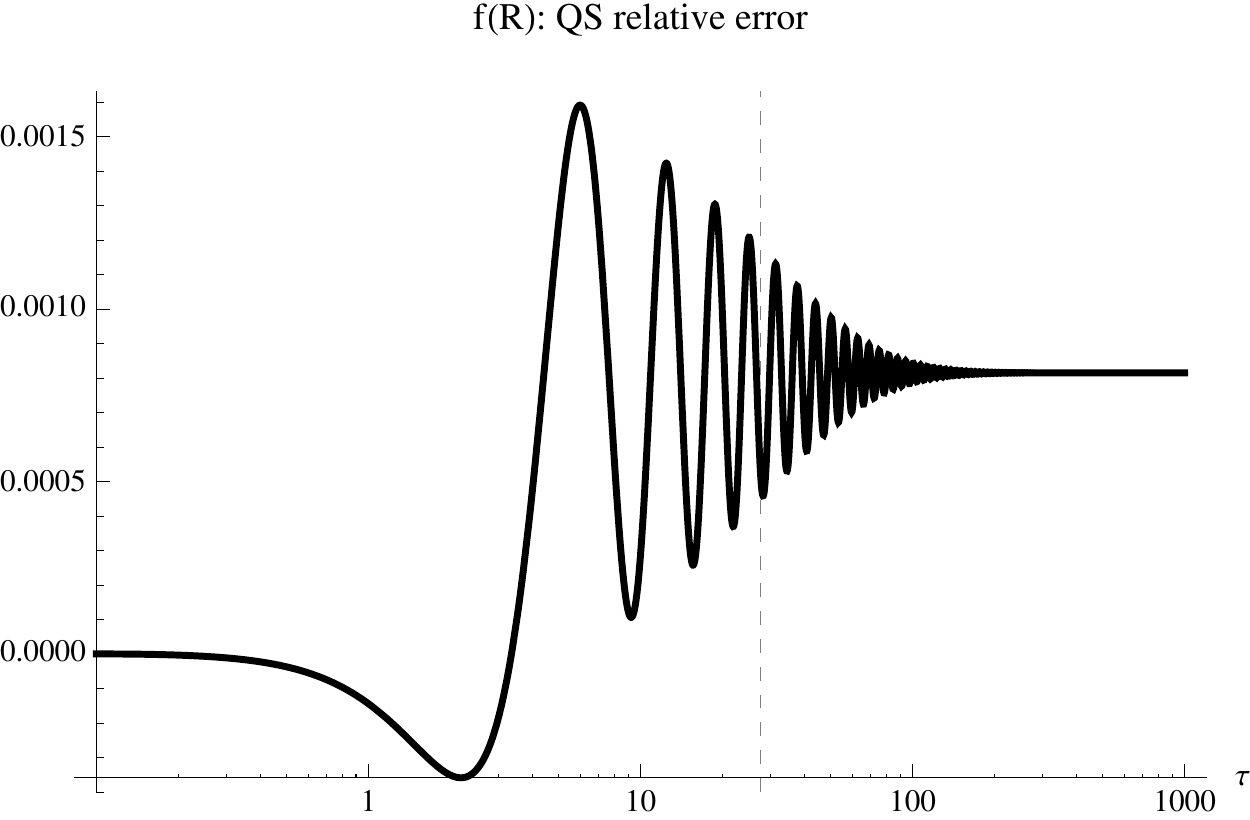}\\
\includegraphics[width=0.5\linewidth]{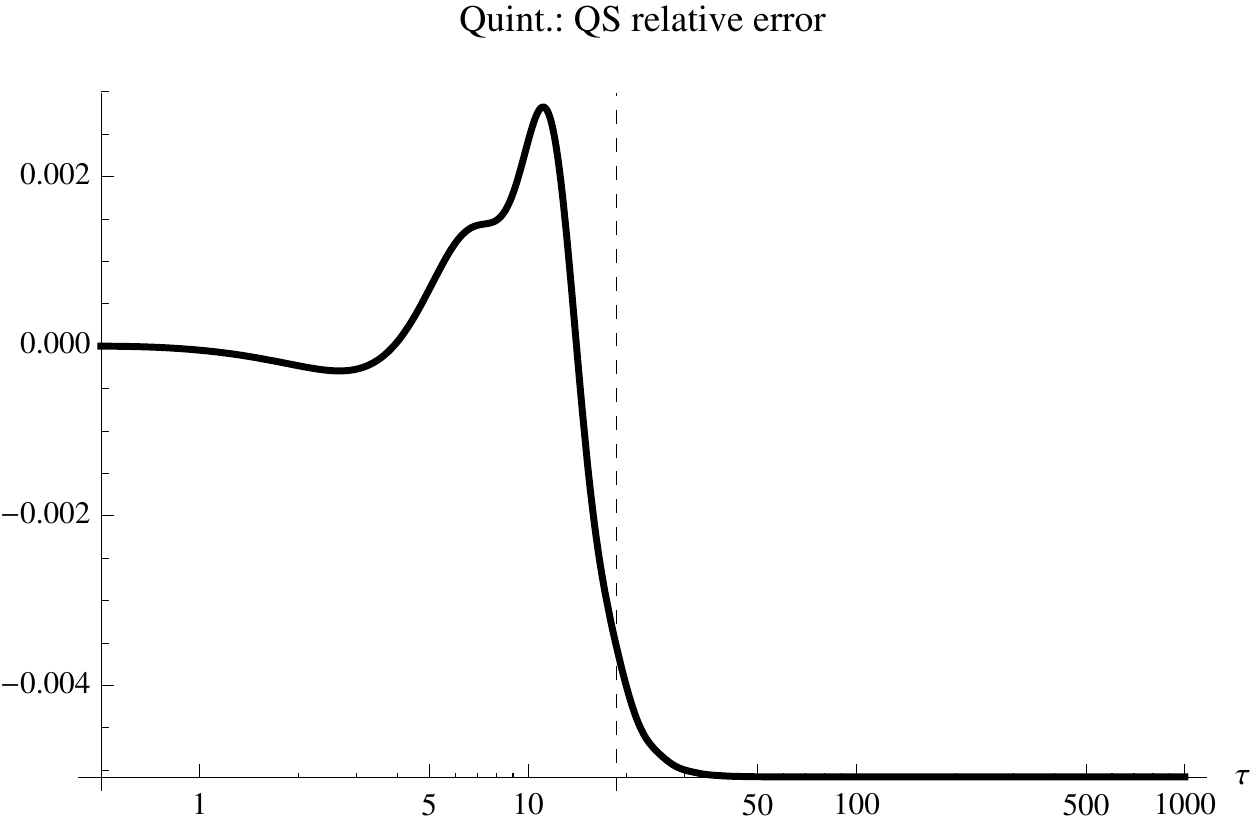}&
\includegraphics[width=0.5\linewidth]{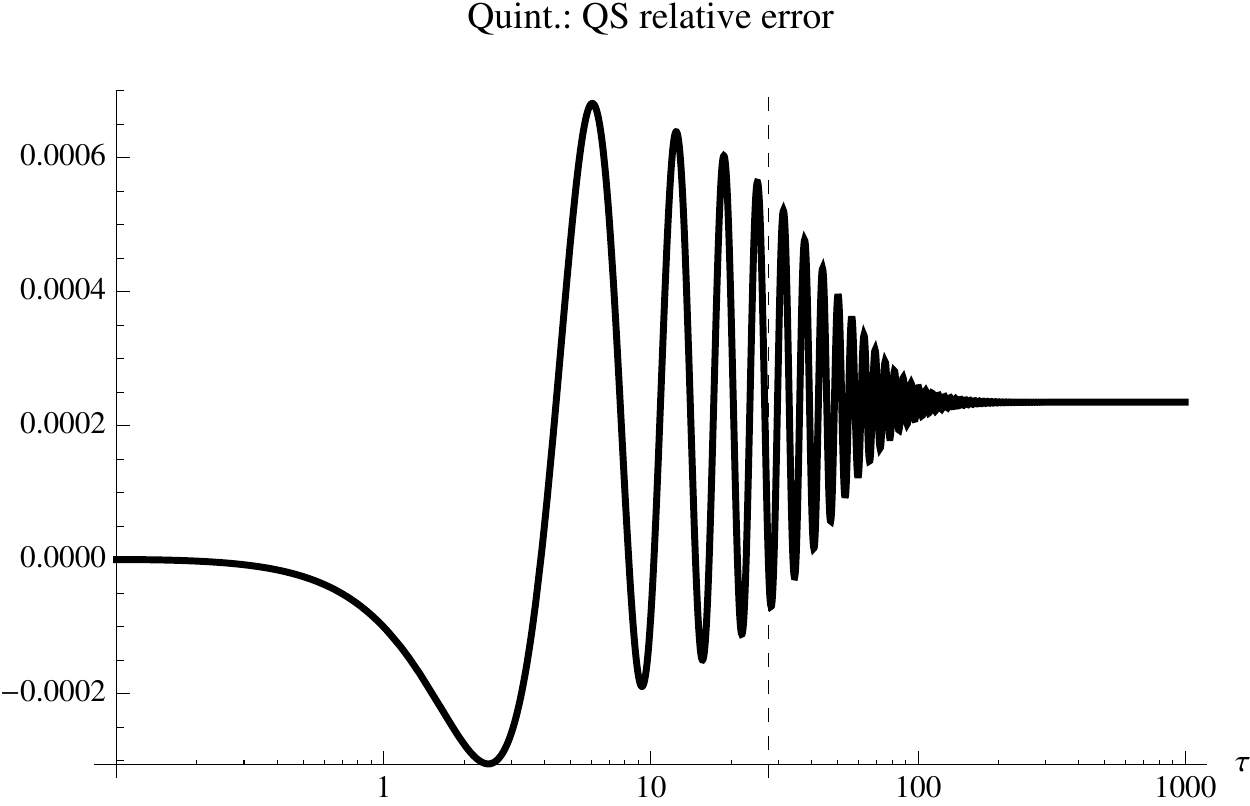}
\end{array}$
\end{center}
\caption{
Plots showing the relative error in $\delta$ (i.e. $\delta_{QS} / \delta_{full} - 1$) again for $f(R)$ (top) and for Quint. (bottom). The fast-rolling case is shown on the left, while the slow-roll case is shown on the right. Note that we have chosen units such that $k = 1$ here, so that $\tau = 1$ corresponds to horizon-crossing and these plots essentially correspond to zooming in on a particular pixel in Figure \ref{fRContour}: $x_{QS} = 0.5$ and $\Omega_\phi (\tau_{QS}) \sim 0.016$ for the fast-roll case and $x_{QS} = 0.1$ and $\Omega_\phi (\tau_{QS}) \sim 0.21$ for the slow-roll case. The oscillatory features clearly visible in the slow-roll case are a direct consequence of $\chi$ displaying decaying oscillatory behaviour on sub-horizon scales, which are not present in the quasi-static solutions.
\label{deltaplots}}
\end{figure*}

We will parametrise the onset of the QSA by two variables. Firstly $x_{QS} = k \tau_{QS}$, labelling the \lq{}time\rq{} when the QSA is switched on. If $x_{QS} > 1$ we are in the sub-horizon regime, whereas $x_{QS} < 1$ indicates the super-horizon regime where we would na\"{i}vely expect the QSA to fail. Secondly, we keep track of the value of $\Omega_\phi$ at  the corresponding time $\tau_{QS}$. We expect this to be relevant, because for a given matching time $x_{QS}$, the QSA should do better the less-dominant the scalar field is. This is because inaccuracies in the evolution of $\chi$ introduced by the QSA should be less consequential for the evolution of $\delta$. Even though the QSA is only designed to hold for sub-horizon times $x_{QS} \gg 1$, it may therefore still be possible that it faithfully reproduces the full evolution on larger scales. In general, however, we expect the following broad features: for large $\Omega_\phi$ and small $x_{QS}$ we should generate large errors, whereas for small $\Omega_\phi$ and large $x_{QS}$ the QSA should be an excellent approximation. 

A few further remarks are in order before proceeding with the QSA analysis for our $f(R)$ scenarios. For the $f(R)$ case we can define the effective potential
\begin{equation} \label{effpot}
 V_{\text{eff},\phi} = V_{,\phi} - \frac{1}{2}\beta\xtilde{\rho}_m
\end{equation}
in terms of which we can also look at the effective equation of state for the scalar degree of freedom
\begin{equation}
w_{\text{eff}} = \frac{1/2 \dot\phi^2 - V_{\text{eff}}}{1/2 \dot\phi^2 + V_{\text{eff}}}.
\end{equation}
A slow-rolling model with $\dot\phi^2 \ll V_{\text{eff}}$ therefore automatically means the scalar field mimics a $\Lambda$CDM evolution with $w \sim -1$ very well.  Fast-rolling solutions will tend to take the background away from $\Lambda$CDM-like behaviour. We may now recall that \cite{0802.2999} found $\Lambda$CDM-like background behaviour to coincide with good quasi-static behaviour in $f(R)$ models on sub-horizon scales. We are now in a position to better understand and quantify why this is the case and also to understand how/whether this statement can be extended to super-horizon scales at all. 

The coefficients of $\chi$ and $\dot\chi$ in Equation \eqref{QSFR}, that determine how much of the QSA error is propagated to the $\delta$ equation respectively are
\begin{eqnarray}
{\cal C}_{\chi} &=& \frac{3\beta}{4} \xtilde{\cal H}^2 \xtilde{\Omega}_m + \xtilde{a}^2 V_{,\phi},  \label{chico} \\
{\cal C}_{\dot\chi} &=& -2\dot\phi. \label{chidotco}
\end{eqnarray}
The second coefficient is clearly suppressed in the $\Lambda$CDM-like slow-roll case when $\dot\phi \ll 1$. The first coefficient can be re-expressed as
\begin{equation}
{\cal C}_{\chi} = \frac{3\beta}{4} \xtilde{\cal H}^2 \xtilde{\Omega}_m + \xtilde{a}^2 V_{,\phi} = \frac{\beta}{4} \xtilde{a}^2 \xtilde{\rho}_m + \xtilde{a}^2 V_{,\phi}.
\end{equation}
It is less obvious how this coefficient will be related to fast- and slow-roll behaviour, so we will investigate this in more detail below. 

Above we have already specified that we will use a fiducial potential $V \sim \Exp{-\abs{\lambda}\phi}$ as studied by \cite{FerreiraJoyce,0611321}. From our expression for the effective potential \eqnref{effpot} we can see that this always has a negative gradient and consequently is a runaway effective potential without a minimum. In the next section we will discuss what happens when the effective potential displays a minimum (the chameleon case). But for now it suffices to notice that with a choice of potential $V \sim \exp{(-\abs{\lambda}\phi)}$, both $V$ and the $\beta$-dependent contribution to the effective potential display runaway behaviour in the same direction
\begin{equation}
V_{\text{eff},\phi} = -|\lambda| V - \frac{\beta}{2} \xtilde{\rho}_m.
\end{equation}
As a direct consequence the $f(R) (\beta = \sqrt{2/3})$ case will have a steeper potential than the corresponding ($\beta = 0$) Quintessence model. This makes slow-roll solutions harder to come by in this particular $f(R)$ model.

\subsection*{Fast Roll}
First we consider an evolution where $\dot\phi$ swiftly becomes non-negligible, \ie{} the field is rolling quickly\footnote{The initial conditions chosen are: $\phi_i = 5, \dot\phi_i = 0, a_i = 1, \lambda = 1.5, \tau_i = 10^{-3}$ and $\xtilde{\rho}_{m}^{*0} \simeq 10$ for $f(R)$ while $\rho_i = \xtilde{\rho}_{m}^{*0} e^{-\beta/2 \chi_i}$ for Quint., so that $\Omega_{\phi,i}$ is identical for the $f(R)$ and Quintessence models. The initial conditions result in a very small ($\sim 10^{-4}$) initial $\Omega_\phi$.}.
The evolution of $\Omega_\phi$ is shown in the left graph of \figref{OmegaCompare}. We compare it to a corresponding non-scaling (Quint) Quintessence model (\ie{} same potential with $\beta = 0$), where the initial condition $\phi_i$ has been chosen so that $\Omega_{\phi}(\tau_{\text{initial}})$ is identical for both cases.
The QSA contour plot for this case is shown in the left graph of \figref{fRContour}. We plot the relative error $\delta_{QS}/\delta_{full} - 1$ to show how well the QSA does in comparison with the full linearised solution. We cut off the evolution and evaluate errors when $\Omega_\phi = 0.7$, \ie{} our model resembles the state of the universe today\footnote{In an explicit N-body context one may want to refine this to only extend to the time where a given scale of interest starts to display non-linear behaviour.}. As explained above we plot the final relative error in the parameter space specified by $x_{QS} = k \tau_{QS}$, the \lq{}time\rq{} when the QSA was switched on, and the value of $\Omega_\phi$ at $\tau_{QS}$. 

A notable feature of \figref{fRContour} is that the error eventually decreases for large values of $\Omega_\phi$. Note that this is an artefact of cutting off the evolution of the error as soon as an $\Omega_{\phi \text{, final}} = 0.7$ is reached. Consequently, if the quasi-static approximation is only switched on at a time when, say, $\Omega_\phi = 0.5$, then even though the QSA will get the evolution of $\delta$ very wrong for super-horizon scales, there is just not very much time left until $\Omega_{\phi \text{, final}} = 0.7$ is reached, so there is very little time for the error to grow. If a different cutoff at an asymptotic value of $\Omega_{\phi \text{, final}} \to 1$ was chosen, and we proceeded to make the analogous contour plot, the error would no longer eventually decreases for large values of $\Omega_\phi$.
Also note that, since $\Omega_\phi$ is still evolving significantly towards its asymptote $\Omega_\phi \to 1$ when the snapshot that leads to \figref{fRContour} is taken (\ie{} when $\Omega_\phi = 0.7$), this means the error can also still be evolving.  This is demonstrated by comparing Figures \ref{OmegaCompare},\ref{chicoeff} and \ref{deltaplots}. The overall error-levels plotted in \figref{fRContour} can therefore continue to grow if a larger $\Omega_{\phi\text{, final}}$ is chosen.

The behaviour of the \qsa{} for the fast-roll case matches our na\"{i}ve hypothesis.  On sub-horizon scales it performs well irrespective of the initial conditions or the model considered, whereas on super-horizon scales the $f(R)$ model does significantly worse than its Quint. counterpart. To see why, we recall that errors in the QSA for $\delta$ stem from propagating an incorrect solution for $\chi$. So we need to investigate how this error propagates to the evolution equation for $\delta$ - in other words, check the coefficients of both $\chi$ as well as $\dot\chi$ in the $\delta$ evolution equation. These are purely {\it background quantities}. They are shown in the two left graphs of \figref{chicoeff} and one can immediately read off the reason why the Quint. model performs significantly better in the QSA than the corresponding $f(R)$ setup.
We can see that the dependence on both $\chi$ and $\dot\chi$ is highly suppressed at early times (i.e. when relevant modes can still be on super-horizon scales) in the Quint. model, explaining why the error in those quantities does not propagate very much at all to the evolution of $\delta$ on those scales. The coefficients plotted in the left graphs of \figref{chicoeff} only become relevant for Quint. at late times, when modes of interest are on sub-horizon scales and where the associated $\chi$ is very well described by its QSA solution. Note that $\Omega_\phi$ also starts evolving later in the Quint. case (as shown in the left graph of \figref{OmegaCompare}), since $\dot\phi \ll 1$ for longer here.

For the $f(R)$ case, on the other hand, we can discern two effects. Firstly the new $\beta$-dependent terms in the evolution equations result in a significant $\chi$-dependence at early times, when $\rho_m$ is still relevant. Secondly, $\dot\phi$ (the coefficient of $\dot\chi$) now also evolves at early times, creating yet another source for the propagation of errors in $\chi$ on super-horizon scales for modes of interest. 
%

The left hand graphs in \figref{deltaplots} finally confirm the intuition gained from the previous plots in this section.
Here we zoom in on a particular case, setting $k=1$, $x_{QS} = 0.5$. This corresponds to a single pixel in the left graph in \figref{fRContour}, namely the pixel at $x_{QS} = 0.5$ and $\Omega_\phi (\tau_{QS}) \sim 0.016$ at the very bottom of the graph: \ie{} this is a point for which the QSA does fairly well. We find that the relative error for the fast-rolling $f(R)$ setup here is approximately an order of magnitude larger than that for the corresponding Quint. model. Finally it may be worth stressing that, while in the fast-roll case the QSA performs badly on scales close to or above the horizon scale, it still performs well on sub-horizon scales as witnessed by \figref{fRContour}, despite having a background evolution that does not closely resemble $\Lambda$CDM (cf. \figref{OmegaCompare}).

\begin{figure*}[htbp] 
\begin{center}$
\begin{array}{ccc}
\includegraphics[width=0.33\linewidth]{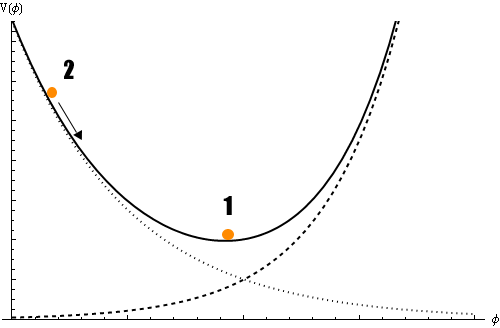} &
\includegraphics[width=0.33\linewidth]{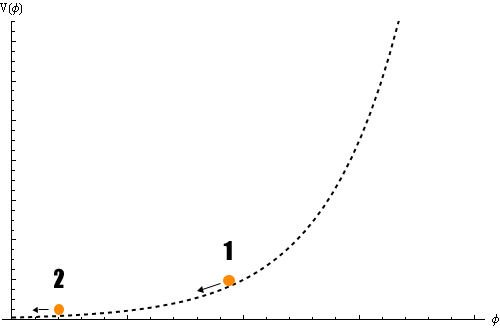}&
\includegraphics[width=0.33\linewidth]{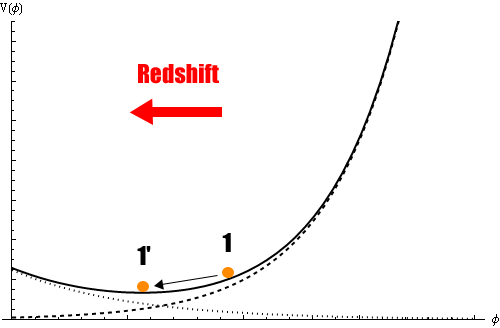}
\end{array}$
\end{center}
\caption{
Here we show the effective chameleon potential and its evolution. 1 and 2 label the two initial conditions for the field $\phi$ considered in the main text. \textsc{Left:} The effective chameleon potential $V_{\text{eff}}$ showing the contributions from the original potential $V(\phi)$ (dashed) and from the non-minimal coupling to matter (dotted).  \textsc{Centre:} The corresponding Quintessence potential, which only possesses the contribution from $V(\phi)$ (dashed) since matter is coupled minimally to gravity.  \textsc{Right:}  Plot showing how the minimum of the effective chameleon potential changes due to the redshifting of the matter-dependent contribution (dotted).
\label{champotplots}}
\end{figure*}
\begin{figure*}[htbp] 
\begin{center}$
\begin{array}{cc}
\includegraphics[width=0.5\linewidth]{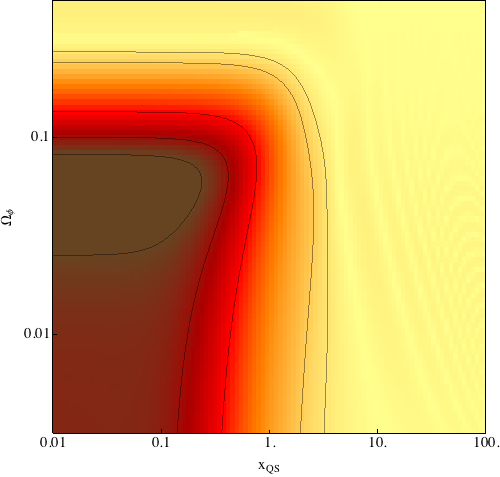} &
\includegraphics[width=0.5\linewidth]{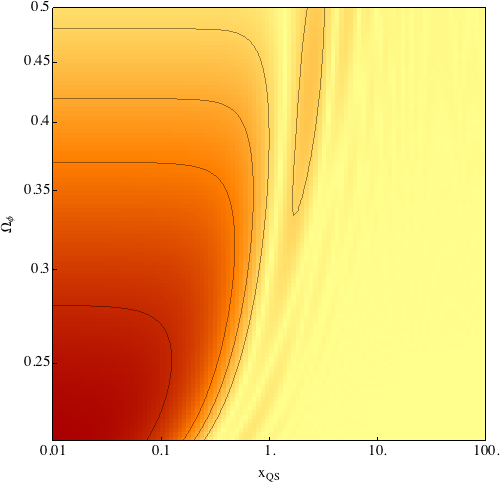} 
\end{array}$
\end{center}
\begin{flushright}$
\begin{array}{cccc}
\includegraphics[width=0.04\linewidth]{whitespace.png} &
\includegraphics[width=0.45\linewidth]{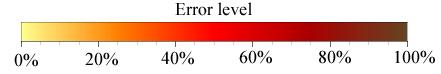} &
\includegraphics[width=0.035\linewidth]{whitespace.png} &
\includegraphics[width=0.45\linewidth]{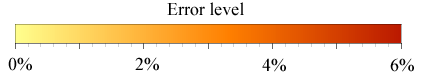}
\end{array}$
\end{flushright}
\caption{
Contour plots plotting the relative error $\delta_{QS}/\delta_{full} - 1$ showing how well the chameleon does in the QSA for the fast-roll initial condition (case 2) away from the minimum on the left and the slow-roll initial condition (case 1) at the minimum of the effective potential on the right  (cf. figure \ref{champotplots}). Note how the slow-rolling nature of the field enforced by case 1 results in a much improved performance of the QSA. Axes are labelled and chosen as in Figure \ref{fRContour} and error contours are $5,10,50,80,100\%$ and $1,2,3,5\%$ from right to left in the fast- and slow-roll cases respectively. The oscillatory features that are visible on (sub-)horizon scales are a consequence of the oscillating behaviour of $\chi$ on those scales, cf. Figure \ref{deltaplots}.
\label{chamcontplots}}
\end{figure*}

\subsection*{Slow Roll} 
Let us now consider a setup with a long initial slow-rolling phase for $\phi$, \ie{} $\dot \phi \ll 1$\footnote{The initial conditions chosen this time are: $\phi_i = 5, \dot\phi_i = 0, a_i = 1, \lambda = 1.5, \tau_i = 10^{-3}$ and $\xtilde{\rho}_{m}^{*0} \simeq 0.016$ for $f(R)$ while $\rho_i = \xtilde{\rho}_{m}^{*0} e^{-\beta/2 \chi_i}$ for Quint., so that $\Omega_{\phi,i}$ is identical for the $f(R)$ and Quint. models.
These initial conditions enforce a relatively large ($\sim 0.2$) initial $\Omega_\phi$ which remains frozen in for a significant amount of time. For contour plots \ref{fRContour} we again evolve forwards until $\Omega_\phi = 0.7$.}.
The evolution of $\Omega_\phi$ in this case is shown in the right graph of \figref{OmegaCompare} and we can immediately spot that the Quint. and $f(R)$ cases behave almost identically. The QSA contour plot for this case is shown in the right graph of \figref{fRContour} and indeed the plot mostly agrees with the corresponding (large $\Omega_\phi$) section of the fast-roll contour plot. However, there is a crucial difference: In the contour plot we show the performance of modes where the QSA is switched on at rescaled time $x_{QS}$ and the background quantity $\Omega_\phi$ is at a given value. But from \figref{OmegaCompare} we know that, due to the initial slow-rolling phase, many modes cross the horizon when $\Omega_\phi$ is still near its initial value. What at first sight might appear to be a numerical artefact in the right graph of \figref{fRContour} --- the fact that there is a very thin strip directly on top of the x-axis (corresponding to the lowest and initial value of $\Omega_\phi$ which happens to be $\sim 0.21$ here and which, during the initial phase of the evolution, remains frozen-in as shown in Figure \ref{OmegaCompare}) and that the QSA does in fact do very well even for modes crossing the horizon during this initial phase --- is a direct consequence of the slow-rolling behaviour of the solution. 

This may appear counter-intuitive, since a large $\Omega_\phi$ means the scalar field is relevant to the cosmic evolution and should hence affect $\delta$. By introducing errors into the evolution of $\chi$ via the QSA, these should then map onto significant errors for $\delta$.  
However, we have already seen above that it is in fact other background properties --- such as the slow- or fast-rolling nature of $\dot\phi$ --- that control how much the QSA errors in $\chi$ are propagated to the evolution of $\delta$. To make this clear let us once again zoom in on a particular case, setting $k=1$ and $x_{QS} = 0.1$. This corresponds to a single pixel in the right graph in \figref{fRContour}, this time the pixel at $x_{QS} = 0.1$ and $\Omega_\phi = 0.21$ in the thin bright (i.e. low error) strip directly at the bottom of the graph; a point for which the QSA does very well as depicted in Figure \ref{deltaplots}. 


As before, we now need to check whether the error introduced into $\chi$ is enhanced or suppressed by the background coefficients in the $\delta$ evolution equation.
These are shown in the right hand graphs in \figref{chicoeff}. Comparing with the corresponding Quint. graphs we see that the background behaviour enforces small coefficients ${\cal C}_{\chi}$ and ${\cal C}_{\dot\chi}$, suppressing the dependence on $\chi$ of the evolution equation for $\delta$ at early times both for the $f(R)$ and Quint. cases this time. For the modes of interest (subhorizon today) the relevant coefficients only become large after horizon-crossing when the exact and QSA solutions for $\chi$ match very well. This is a consequence of the initial slow-rolling phase.
The conclusion one draws here is that, once the evolution equations for the perturbations are known, we can understand how well the QSA performs on super-horizon scales in terms of background quantities.  In the particular case considered here, even though we started with a large $\Omega_\phi$, this remained frozen in initially so that $\dot\phi$ remained small and the dependence on $\chi$ is also suppressed. 
The right hand graphs in \figref{deltaplots} summarise these results, showing that the relative errors for both the $f(R)$ and Quint. setups considered in this section are very small (on the sub $0.1\%$ level).

The key result of this section is that the impact of the QSA can depend crucially on how the evolution equation for the scalar field couples back into that of the density perturbation. Small errors in the QSA for $\chi$ can be greatly amplified if the background scalar field evolves substantially. 
%
Small values of $\Omega_\phi$ (indicating that the field $\phi$ only negligibly contributes to the energy density of the universe at the relevant time) may not be enough to prevent the propagation of large errors. In some sense, this is not surprising- it is the {\it non-static} nature of the background which is pushing the QSA outside its range of validity. And, if the QSA is to be applied in any specific $f(R)$ theory, it is clearly essential to check whether the evolution of the scalar field is such that the approximation is good enough.

\section{The Chameleon mechanism in $f(R)$}
\label{SCR}
%


\begin{figure*}[tp] 
\begin{center}$
\begin{array}{cc}
\includegraphics[width=0.5\linewidth]{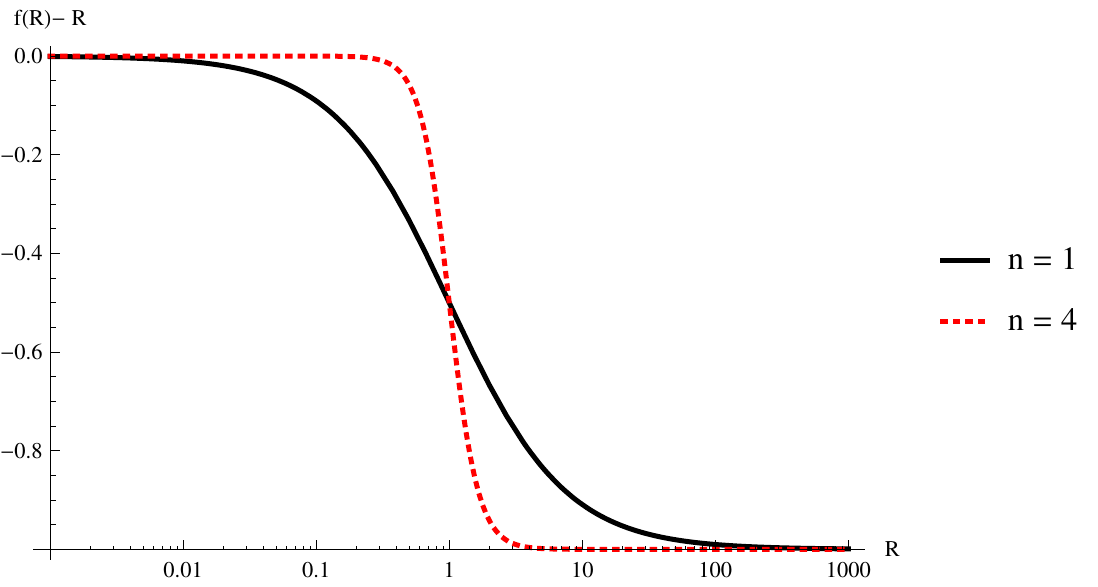} &
\includegraphics[width=0.5\linewidth]{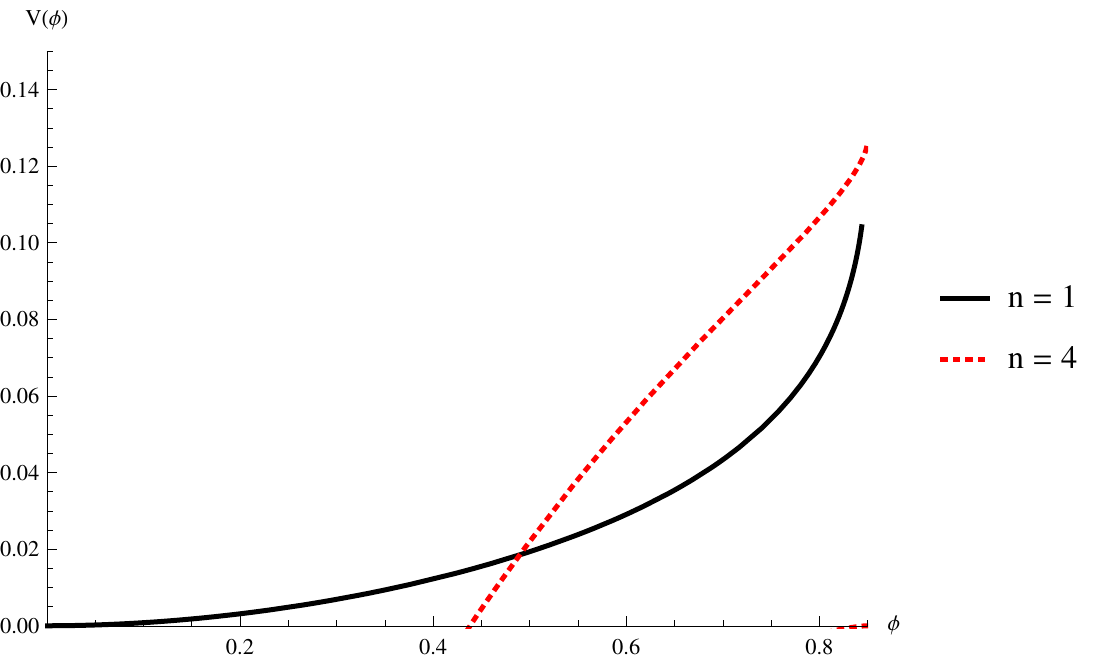}
\end{array}$
\end{center}
\caption{
The Hu \& Sawicki potential (equation \eqref{HSpot}) for $m=c_1=c_2=1$. \textsc{Left}: We plot $f(R) - R$ vs. $R$, showing how this model interpolates between different $f(R)$ for large and small curvatures. \textsc{Right}:  The resulting $V(\phi)$. Note how the potential for $n=1$ satisfies $V_\phi, V_{\phi\phi},V_{\phi\phi\phi} >0$ for all $R$ (and hence always acts as a chameleon), whereas $n=4$ only satisfies this for large $\phi \sim 0.8$ (which corresponds to large curvature $R$ here), so chameleon-like behaviour is restricted to the high curvature regime in the second case.
\label{fig-HSpotentials}}
\end{figure*}

\begin{figure*}[htbp] 
\begin{center}$
\begin{array}{cc}
\includegraphics[width=0.5\linewidth]{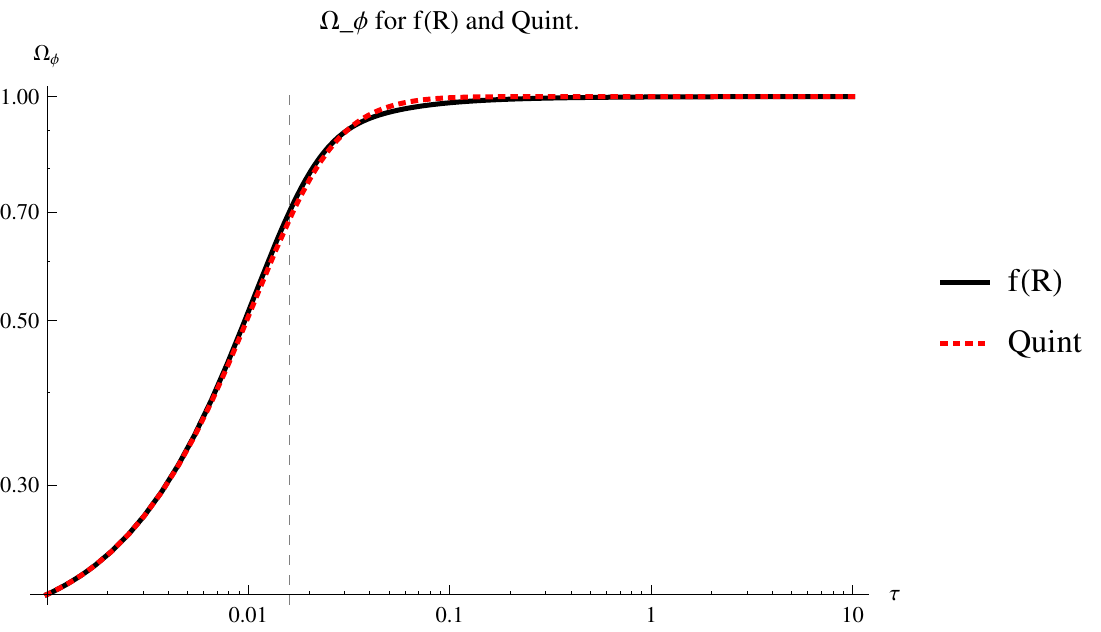} &
\includegraphics[width=0.5\linewidth]{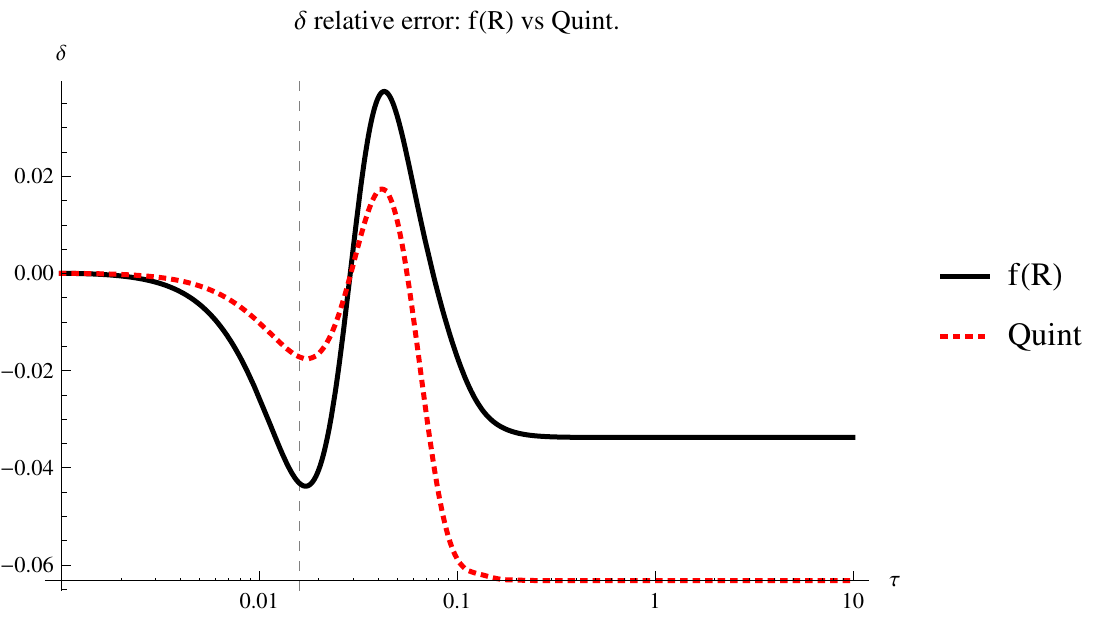}\\
\includegraphics[width=0.5\linewidth]{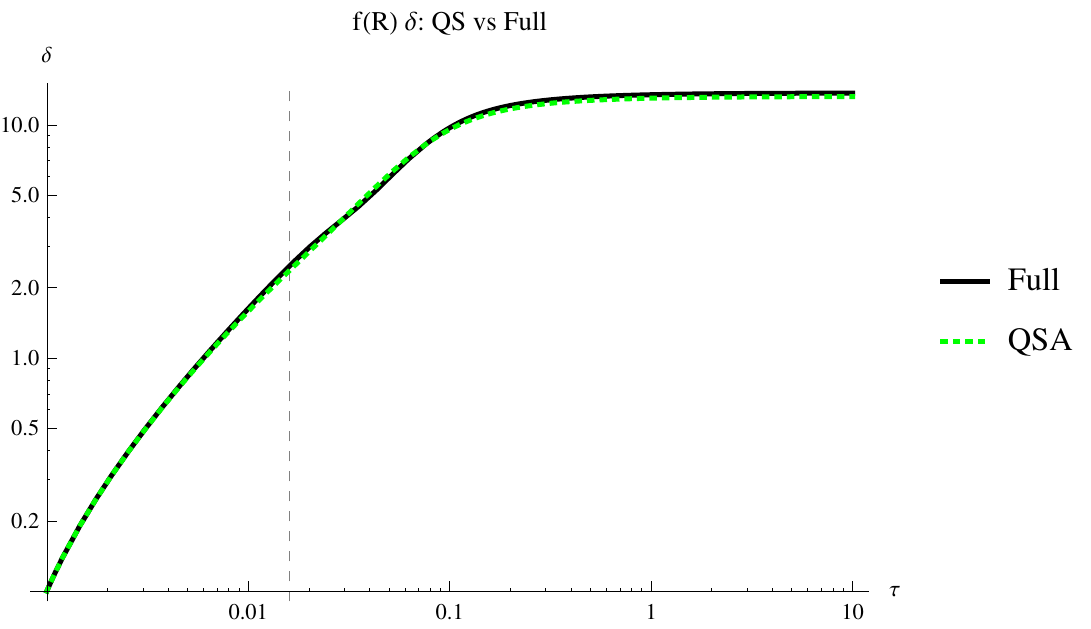}& 
\includegraphics[width=0.5\linewidth]{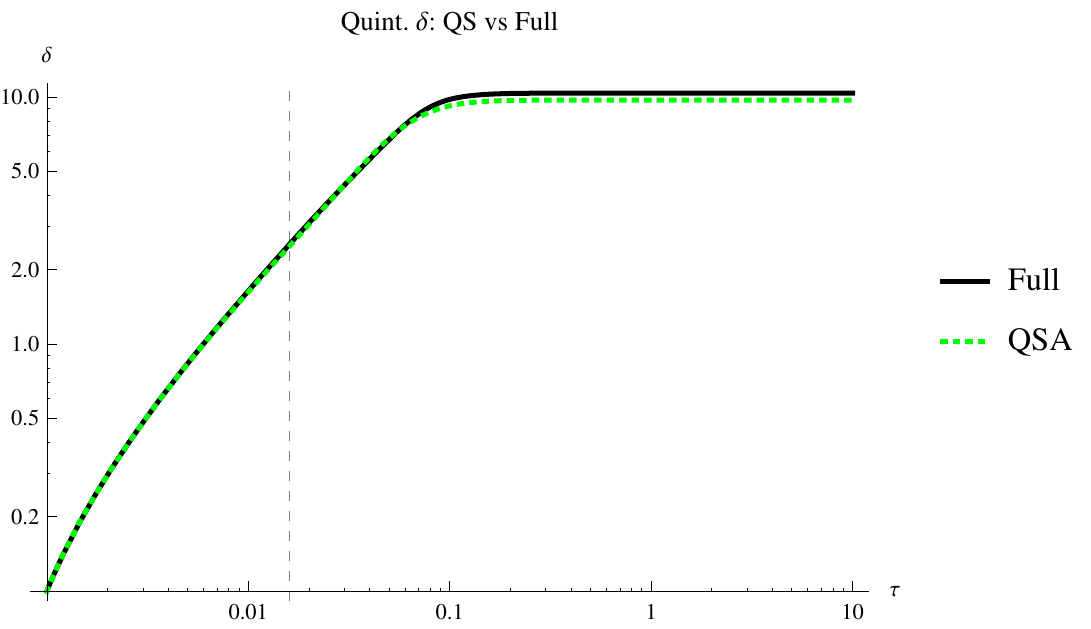}\\
\includegraphics[width=0.5\linewidth]{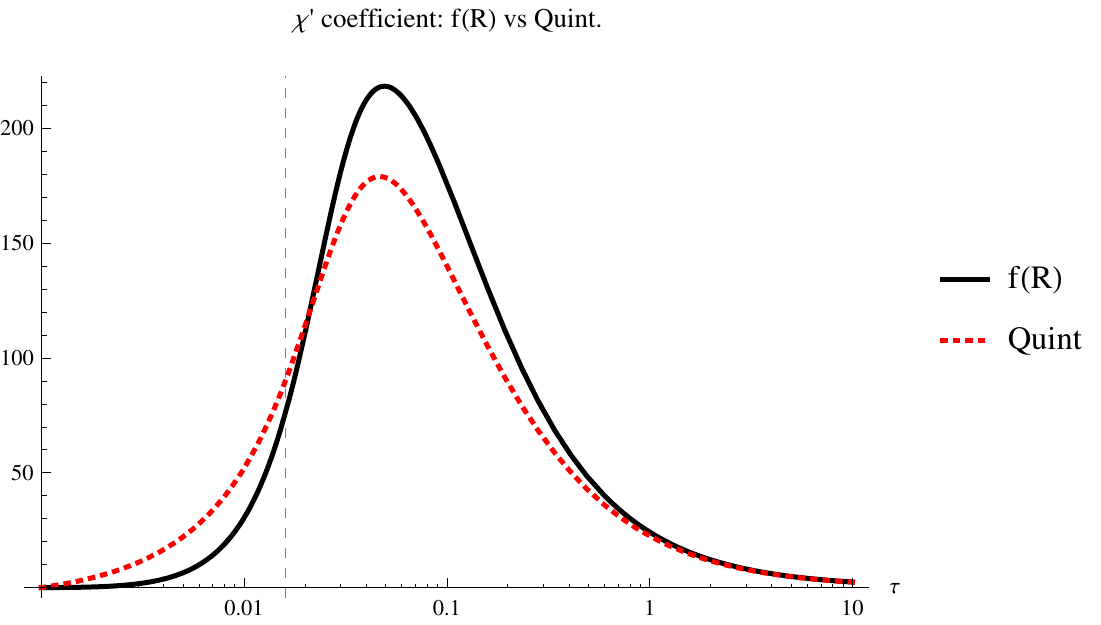}& 
\includegraphics[width=0.5\linewidth]{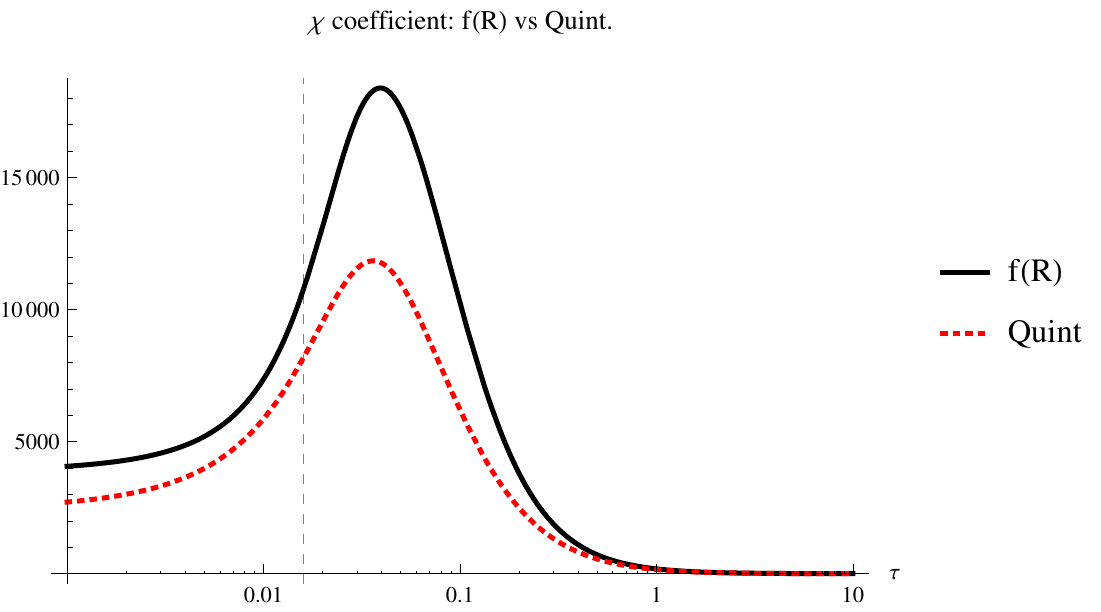}
\end{array}$
\end{center}
\caption{
The slow-rolling chameleon (case 1): Initial conditions place the field at the minimum in the effective potential $V_{eff}$, resulting in a slow-rolling field and small QSA errors. The Quintessence-like case also performs well due to the very flat $V(\phi)$. The dashed horizontal line denotes the time when $\Omega_\phi = 0.7$ and the relative error is evaluated in our contour plots. \textsc{Top row}: We show the evolution of $\Omega_\phi$ for a chameleon $f(R)$ and Quint. model starting with identical $\Omega_\phi$ on the left. Note these evolutions are almost identical. On the right we show the evolution of $\delta_{QS}/\delta_{full} - 1$ in units where $k=100$ and choosing $x_{QS} = 0.1$ and $\Omega_\phi (\tau_{QS}) \sim 0.22$ (cf. Figure \ref{chamcontplots}. Horizon-crossing therefore takes place at $\tau = 0.01$. \textsc{Middle row}: The evolution of $\delta$ in the $f(R)$ chameleon case on the left and the Quint. case on the right, showing full and quasi-static solutions, which agree very well in the slow-roll case shown here. \textsc{Bottom row}:  Evolution of coefficients for $\dot{\chi}$ (left) and $\chi$ (right) in \eqref{QSFR} - note that chameleon $f(R)$ and Quint solution closely follow each other here in comparison to the analogous plots in Figure \ref{deltaplots-ChamMil} (up to $\sim 50\%$ vs. $> 1000\%$ difference).
\label{deltaplots-ChamMin}}
\end{figure*}

It is well-known \cite{KhouryWeltman,Brax:2008hh} that a subset of $f(R)$ models give rise to the so-called chameleon effect, where the non-minimal coupling to matter in the Einstein frame results in an effective potential for $\phi$ with a minimum, and consequently an effective mass. In chameleon models this is used to screen away any fifth force from $\phi$ in dense regions, allowing them to evade tight fifth force constraints on solar system scales \cite{KhouryWeltman}. Such a screening mechanism is therefore an essential ingredient to construct an observationally viable $f(R)$ model. Screening is an intrinsically non-linear effect and our linearised analysis is consequently not sensitive to it by default. However, the analysis is sensitive to the form of the potential via the associated mass term\footnote{After all, the background field evolution and especially $\dot\phi$ are highly sensitive to the form of the potential.}, so it is worth considering how this impacts our analysis and whether there are any interesting consequences for the QSA. 

The $f(R)$ model considered in the previous section does not display chameleonic behaviour. This is straightforward to understand from the background evolution equation. Recall this is 
\begin{equation} \label{backgroundphieom}
\ddot\phi + 2 \xtilde{\hub}\dot \phi + \xtilde{a}^{2} V_{\phi}= \frac{1}{2}\beta \xtilde{a}^{2} \xtilde{\rho}_{m}  
\end{equation}
for the background scalar $\phi$. Now we can write this in terms of an effective potential for $\phi$ (absorbing the factor $\xtilde{a}^2$ this time)
\begin{equation}
V_{\text{eff}, \phi} =  \xtilde{a}^{2} V_{\phi} - \frac{1}{2}\beta \xtilde{a}^{2} \xtilde{\rho}_{m}.  
\end{equation}
However, for the runaway potential $V \sim e^{-|\lambda| \phi}$ both contributions to $V_{\text{eff}, \phi} $ are negative, so no minimum exists. Yet, for a chameleon-like model, we require that $V_{\text{eff}}$ has a minimum. 

Under what conditions does the $f(R)$ potential fulfil the requirements for chameleon behaviour?  Adapting the results of \cite{Brax:2008hh} to the conventions used throughout this paper, we find that the relevant conditions are\footnote{This may come as a surprise, given the result of \cite{Brax:2008hh} who quote the condition: $V_\phi < 0  \; V_{\phi\phi} > 0 \; V_{\phi\phi\phi} < 0$ as required for $f(R)$ models with chameleon screening. This difference is due to two differing conventions used in the literature when mapping a given $f(R)$ model into its scalar-tensor form. We discuss these conventions in Appendix \ref{CT} and describe the field redefinition that maps between them. Also note that we need $V_\phi > 0$, since for a minimum we require $V_{eff,\phi} = 0 $, but the contribution from the non-minimal coupling to matter to $V_{eff,\phi}$ is negative.}
\begin{align}
V_\phi &> 0    &V_{\phi\phi} &> 0   &V_{\phi\phi\phi} &> 0.
\end{align}
We can check that this is indeed the case. Firstly consider a new fiducial chameleon potential $V = \Exp{\abs{\lambda} \phi}$ trivially satisfying the chameleon conditions above. From equation \eqref{backgroundphieom} this can clearly generate an extremum for the effective potential now. The derivatives of the potential in our convention are now given by
\begin{align}
V = \; &\frac{R f_R-f}{2 \left(f_R+1\right)^2}\\
V_\phi =\; &\beta  \frac{R + 2 f  -R f_R}{2 \left(1+f_R\right)^2}\\
V_{\phi\phi} = \; &\frac{\beta ^2}{2}  \left(\frac{1}{f_{RR}}+\frac{R \left(f_R-3\right)-4 f}{\left(1+f_R\right)^2}\right)\\
\nn V_{\phi\phi\phi} = &-\beta ^3\frac{f_{RRR} \left(f_R+1\right)^3+3 \left(f_R+1\right)^2 f_{RR}^2}{2 \left(1+f_{R}\right)^2 f_{RR}^3} \\
&-\beta ^3 \frac{R \left(f_R-7\right)-8 f}{2 \left(1+f_{R}\right)^2}.
\end{align}
In order for the effective potential $V_{\text{eff}}$ to have a minimum in the Jordan frame, the condition 
\begin{equation}
R + 2 f  -R f_R > 0
\end{equation}
needs to be satisfied \cite{0705.1158,0611321}. Comparing with our expression for the derivatives of the potential, this shows that $V_\phi > 0$ as expected. As a further check we can cross-check against a model that is known to have consistent chameleon screening, the Hu \& Sawicki model \cite{0705.1158}. Figure \ref{fig-HSpotentials} demonstrates that regions of parameter space satisfy the necessary constraints for different choices of parameters in this model. As a corollary we see that the fiducial exponential potential we have chosen here qualitatively is a good proxy for Hu \& Sawicki potentials in regions of parameter space that display chameleon screening. 

\begin{figure*}[htbp] 
\begin{center}$
\begin{array}{cc}
\includegraphics[width=0.5\linewidth]{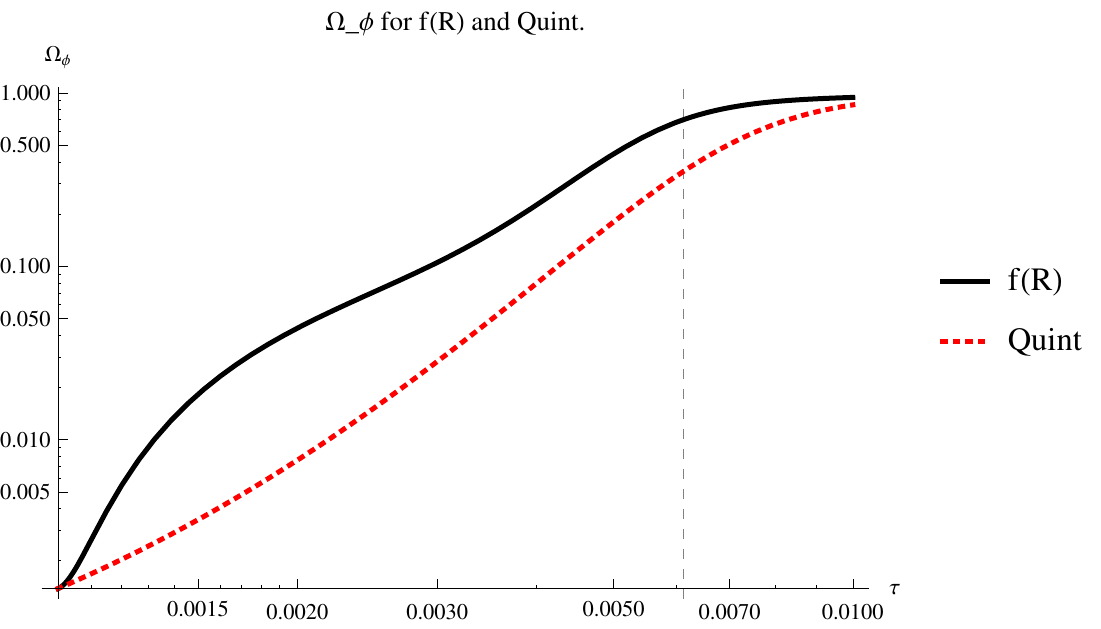} &
\includegraphics[width=0.5\linewidth]{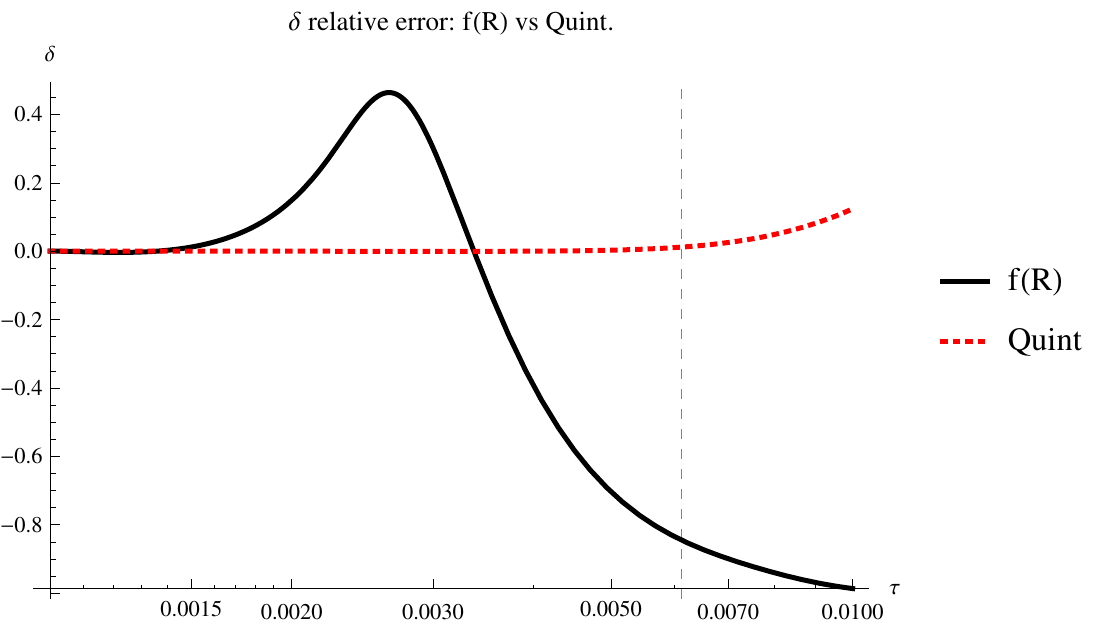}\\
\includegraphics[width=0.5\linewidth]{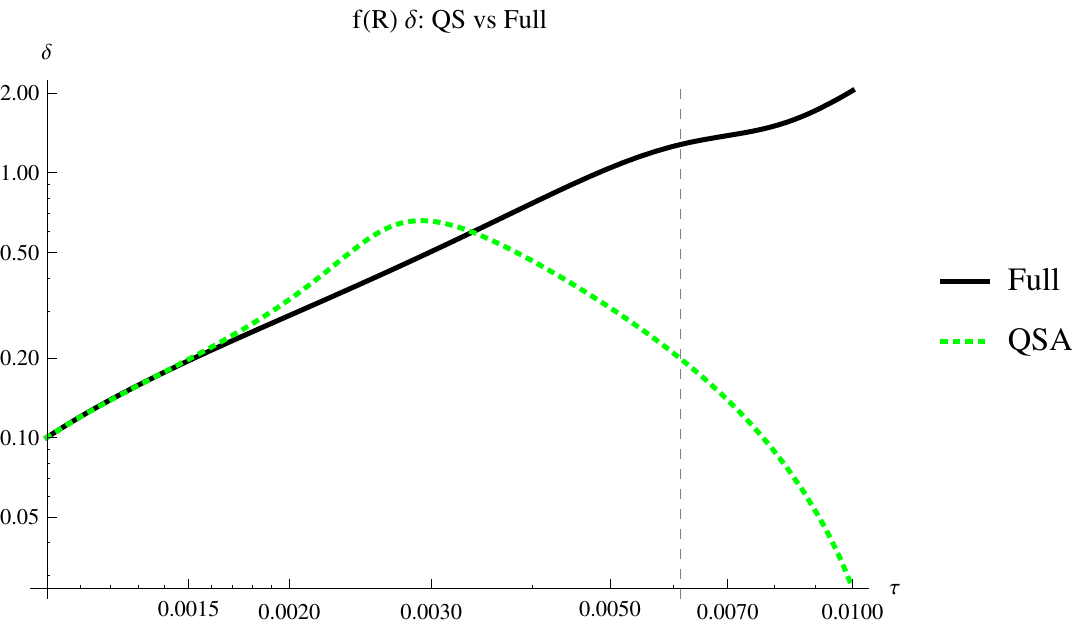}& 
\includegraphics[width=0.5\linewidth]{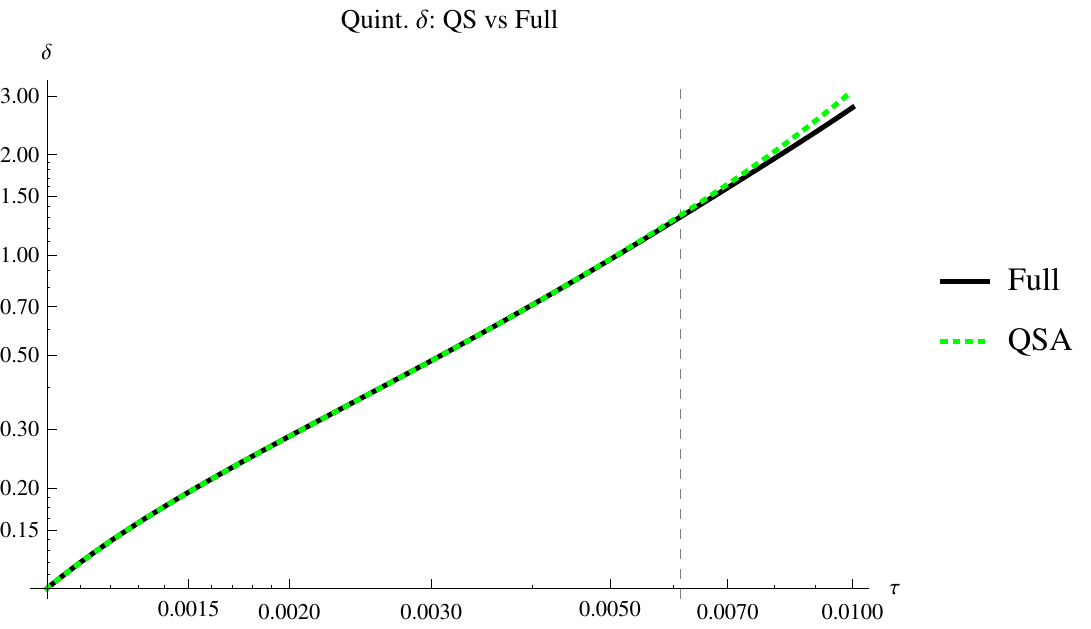}\\
\includegraphics[width=0.5\linewidth]{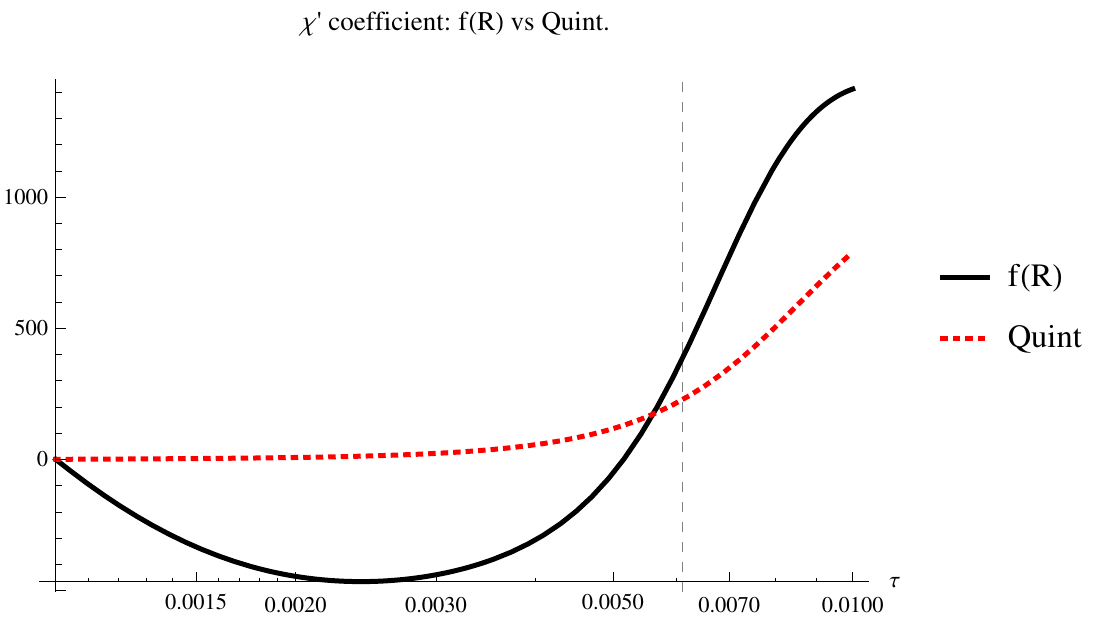}& 
\includegraphics[width=0.5\linewidth]{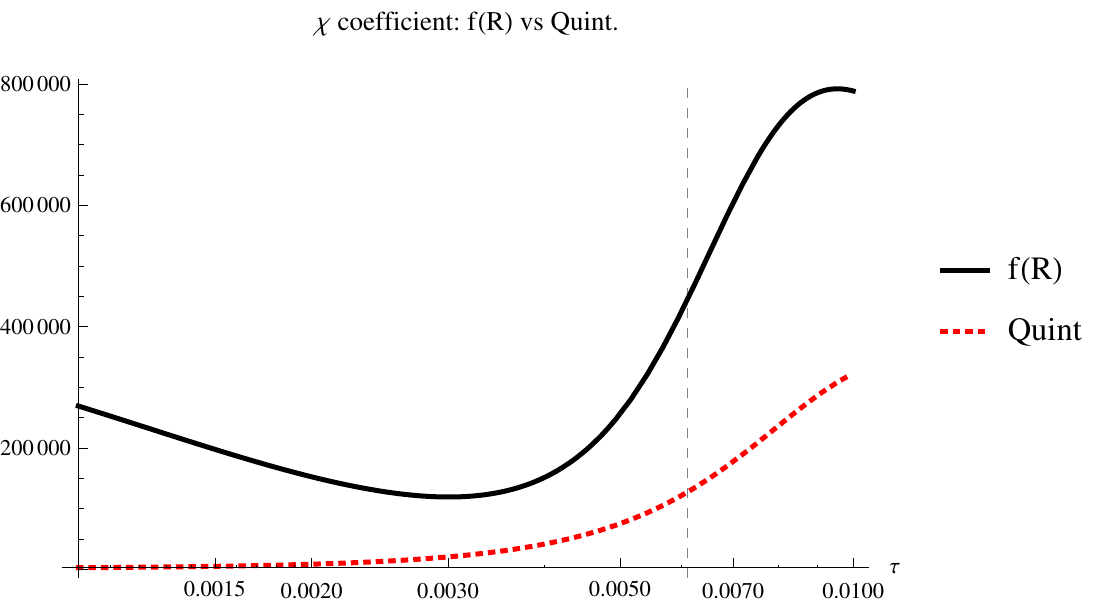}
\end{array}$
\end{center}
\caption{
The fast-rolling chameleon case (case 2): Initial conditions place the field away from the minimum in the $f(R)$ model, resulting in a fast-rolling field and large QSA errors, markedly different from case 1 shown in Figure \ref{deltaplots-ChamMin}. The corresponding Quintessence-like performance is hardly changed in comparison with case 1 as expected. The dashed horizontal line denotes the time when $\Omega_\phi = 0.7$ and the relative error is evaluated in our contour plots. 
\textsc{Top row}: We show the evolution of $\Omega_\phi$ for a chameleon $f(R)$ and Quint. model starting with identical $\Omega_\phi$ on the left. Note these evolutions are visibly different now. On the right we show the evolution of $\delta_{QS}/\delta_{full} - 1$ in units where $k=100$ and choosing $x_{QS} = 0.1$ and $\Omega_\phi (\tau_{QS}) \sim 0.0014$ (cf. Figure \ref{chamcontplots}. Horizon-crossing therefore takes place at $\tau = 0.01$. \textsc{Middle row}: The evolution of $\delta$ in the $f(R)$ chameleon case on the left and the Quint. case on the right, showing full and quasi-static solutions - the QSA fails rather catastrophically in the $f(R)$ chameleon here, while the Quint. QSA solution faithfully tracks the full solution. Again this is in stark contrast to the slow-roll case considered before and is a result of fact that in the fast-roll case there is no suppression of the propagation of large quasi-static errors for $\chi$ to the evolution of $\xtilde{delta}$ on super-horizon scales. \textsc{Bottom row}:  Evolution of coefficients for $\dot{\chi}$ (left) and $\chi$ (right) in \eqref{QSFR} - note that chameleon $f(R)$ and Quint solution are very different now, with the $f(R)$ chameleon displaying much larger coefficients. This explains why the QSA error in evaluating $\chi$ is much more strongly propagated into the evolution equation for $\delta$, resulting in the bad QSA fit shown in the middle row. Contrast this with the case shown in Figure \ref{deltaplots-ChamMin}.
\label{deltaplots-ChamMil}}
\end{figure*}

Equipped with the above insights about the nature of the potential, we choose a fiducial chameleon potential $V = e^{|\lambda| \phi}$. The evolution equations for the background, perturbations and perturbations in the QSA laid out in the previous section are still valid. We now contrast two cases. 
In the first case, we initially place the field at the minimum (this is case 1 shown in the left graph of figure \ref{champotplots}); in this situation we expect the QSA to do very well and indeed it does as shown in the left graph of figure \ref{chamcontplots}. This is because, as we saw in the previous section, errors generated by the QSA are propagated via their dependence on $\dot \phi$ and $\hub^2 \Omega_m$. If the field is slow-rolling any dependence on $\chi$ is highly suppressed; indeed, starting at the minimum should keep $\phi$ frozen at the minimum. 
Having said that, since the effective potential will evolve due to the redshifting of matter density, the field will in fact slowly roll tracking the effective minimum, so a small error should still remain. This effect is shown in the right graph in Figure \ref{champotplots}, while the middle graph in the same figure shows the corresponding situation in the Quint. setup which lacks any contribution to the effective potential that depends on the cosmological matter density (again we match initial conditions so that the Quint. and chameleon cases start off with the same $\Omega_\phi$ as discussed in the previous section).

The initial condition for starting out at the minimum of the potential is
\begin{equation}
V_{\text{eff},\phi} = \sqrt{1/6} \xtilde{\rho}_m - \lambda V_0 e^{\lambda \phi} = 0.
\end{equation}
Denoting the initial value of the scalar field by $\phi_i$, in terms of an initial condition for the matter energy density $\xtilde{\rho}_m$ this means we require
\begin{equation} \label{rhoini}
\xtilde{\rho}_{m\text{, initial}} = \sqrt{6} \lambda e^{\lambda \phi_i}.
\end{equation}
This means that the initial energy density $\Omega_\phi$, which we may write as
\begin{equation}
\Omega_\phi = \frac{1}{1 + \frac{\rho_m}{\rho_\phi}},
\end{equation}
is fixed once we require the field to start at its minimum and specify $\lambda$. 

To understand this better let us once again effectively zoom in on a single pixel in the contour plot, setting $k=100$, $x_{QS} = 0.1$ and $\Omega_\phi (\tau_{QS}) \sim 0.22$. Also setting $\lambda=1.5$ as for the contour plots we obtain the evolution shown in \figref{deltaplots-ChamMin}.\footnote{Again we emphasize that the parameters ($\lambda$, $\phi_i$, etc.) chosen for our examples are intended to give rise to toy models providing an understanding of the QSA. An in-detail comparison with observational constraints on the parameter space of such models is beyond the scope of this paper.} One sees that the background field $\phi$ is indeed very slowly rolling. We compare this with a Quintessence-like model that starts out with the same $\Omega_\phi$. The reason the non-chameleon Quintessence-like model also does relatively well, is that the minimum  of the effective potential lies in a region where the curvature of the original $\phi$ potential is very small (\cf{} the middle graph in \figref{champotplots}). Hence the field is slow-rolling in the Quintessence case too, only doing mildly worse in the long run than the chameleonic $f(R)$ setup.

In the second case we start away from the minimum. This is labelled as case 2 for both the $f(R)$/chameleon and Quintessence cases in \figref{champotplots}. The QSA error introduced now is shown in the left graph in \figref{chamcontplots} and we see that the QSA does significantly worse than in the the first case, where the field started at the minimum of the effective potential.
Zooming in on a pixel with $k=100$, $x_{QS} = 0.1$ and $\Omega_\phi (\tau_{QS}) \sim 0.0014$, we obtain the evolution shown in \figref{deltaplots-ChamMil}. As expected the Quintessence-like model is hardly affected by the change from case 1 to case 2.  In fact it does slightly better than before since we have effectively moved into the flat, tail end of the original potential for $\phi$. However, the $\rho_m$-dependent term in the effective potential for the chameleon case means the field there is rolling down a very steep slope and hence the QSA error is strongly propagated to the $\delta$ evolution equation, resulting in a very bad fit for the QSA (\figref{chamcontplots}).

While the the two cases considered above are extremely useful in understanding what controls the accuracy of the QSA and in particular in stressing the importance of the fast/slow-rolling nature of the background, at this point it is important that an initial condition very close to or identical to case 1 is the observationally motivated case. Firstly note that BBN constraints require the field to have settled into its minimum by the time BBN starts \cite{Brax:2004px}. CMB constraints can also be used to place bounds on the variation of $\phi$ since recombination, giving \cite{Brax:2013yja}
\begin{equation}
\abs{ \Exp{\frac{\beta \Delta\phi}{M_{Pl}}}   -1 } < 0.05
\end{equation}
This ensures that viable chameleon models do well in the QSA in the linearised regime, since as we have seen, the approximation works well if we start close to the minimum of the effective potential (which results in a maximally slow-rolling evolution). This serves as somewhat of an a posteriori justification for the use of the QSA in chameleon models - and we should stress: even on super-horizon scales. Note that this is directly related to the shape of the chameleon potential. Since the field is slow-rolling along with the effective minimum, QSA errors are strongly suppressed. Of course the effective minimum also generates an effective screening mass for $\phi$. Nevertheless we should keep in mind that, while the screening properties of chameleon theories are intrinsically non-linear effects, the fact that the QSA does well here solely relies on the slow-rolling nature of the background. One should therefore not convolute explanations for the efficiency of screening and the accuracy of the QSA in this case.

\section{Discussion}
\label{D}

What have we learned from our analysis of the QSA in linearised $f(R)$, chameleon and, \textit{en passant}, in Quintessence models? The key insight of this paper is that the performance of the perturbative QSA on a given scale in all of these models can be understood in terms of background variables. This allows us to straightforwardly quantify how well a given model does in the QSA and to assess whether this approximation can also be used in super-horizon regimes. In particular the slow- or fast-rolling nature of the background field plays a crucial role in propagating potential quasi-static errors into structure formation, \ie{} $\xtilde{\delta}$. Slow-rolling solutions lead to quasi-static solutions which perform well even outside their na\"{i}ve regimes of validity, \ie{} also on super-horizon scales. 

Slow-rolling solutions also correspond to $\Lambda$CDM-like background evolutions, which \cite{0802.2999} found to be linked to good quasi-static evolution on sub-horizon scales. Phrasing this in terms of slow- and fast-rolling solutions and investigating the evolution equations \eqref{deltafr},\eqref{chifr} and \eqref{CDMeq2} has allowed us to gain a semi-analytical understanding of why this is the case. We have extended the argument to (super-)horizon scales, where slow-rolling solutions are still accurate within $\sim 5\%$ for the chameleon case considered in Section \ref{SCR}. We have also found that on sub-horizon scales the QSA performs well as expected, with $< 1\%$ level errors in $\xtilde{\delta}$. This can even be the case when the field is fast-rolling and the background evolution is consequently distinct from $\Lambda$CDM, as the fast-roll examples in sections \ref{FR} and \ref{SCR} show\footnote{Note that we do not expect this to remain true in general, for example in cases where there are still rapid oscillations of scalar field perturbations deep into the sub-horizon regime. An explicit example is provided by the $R^{0.63}$ case discussed in \cite{0802.2999}, where the QSA fails on sub-horizon scales too. We thank Antonio Maroto for pointing this out to us.}.  Note that we expect the exact error-levels to be sensitive to the precise nature of the potential, so it will be an interesting task for the future to combine the findings of this paper with an exhaustive survey of observationally viable chameleon and $f(R)$ models.


Viable $f(R)$ and chameleon models satisfy two conditions. Firstly, they come equipped with a screening mechanism that avoids clashes with tight fifth force constraints. This screening mechanism relies on an effective potential with a minimum. Secondly, BBN and CMB constraints require the field to be very close to this minimum by the time of BBN/recombination and to have the field subsequently slow-rolling together with the evolving minimum (we recall that the minimum evolves due to the redshifting matter density). Here we have shown that the resulting slow-roll condition on the evolution of the background field is precisely what is required for the QSA to perform well even on (super-)horizon scales. It therefore appears that the QSA is well-suited to analyse structure formation in such modified gravity models for a range of scales - an encouraging conclusion indeed. This is in agreement with (and an extension of) the conclusions of \cite{0802.2999,1210.6880}, who discuss sub- and near-horizon scales, and the analysis presented here sheds some light on why the QSA performs so well in these cases.

However, note that a question of precision remains. QSA errors introduced into the evolution of $\delta$ can still reach $\sim 5\%$ on super-horizon scales, even in the best cases considered in this paper, so that the use of the QSA limits the maximal precision that can be reached in any such analysis of structure formation.  Such an error is still too large if one targets $1\%$ accuracy for the power spectrum of $\delta$ in future experiments.\footnote{Also note that intrinsic N-body simulation systematics, e.g. due to the discretisation of evolution equations, will introduce further errors. It will be an interesting task for the future to establish precisely at what level these errors contribute. We thank Baojiu Li for raising this point.} 
If higher accuracies are desired a more accurate prescription than one employing the QSA will likely be necessary. Also adding a short fast-rolling phase before BBN or maximising the distance the field could have travelled in accordance with constraints would probably worsen the obtained accuracy. This is of crucial importance in the context of N-body simulations, in particular when the QSA is used to set up initial conditions in the linear
regime on (super) horizon scales, which is precisely the regime we have probed here.

Much remains to be done. Probing Vainshtein screening in the same quantitative fashion and exploring whether there are viable dark energy models that are not well described by the QSA (along the lines of \cite{1302.1774,Silvestri:2013ne}) are tasks left for future work. 
For Vainshtein-screened models it could be very interesting to extend the work of 
\cite{Barreira:2012kk,Barreira:2013xea, Barreira:2013eea, Li:2013tda}, exploring the accuracy of the QSA for such models. The fact that the background evolution can be very distinct from $\Lambda$CDM in such models might suggest that the QSA will generically perform rather poorly on superhorizon scales there. However, a detailed analysis may uncover interesting exceptions. Finally the analysis in this paper has focussed on the linear regime relevant to the way initial conditions are set up in N-body simulations and to (super-)horizon scales. An explicit and detailed fully non-linear analysis of the QSA on sub-horizon scales should also result in a better understanding of the applicability of the approximation.

\begin{acknowledgments}
We thank Sigurd N\ae{}ss, Luca Amendola, Kazuya Koyama, Claudio Llinares, David Mota, Dmitry Pogosyan and Ignacy Sawicki for very useful discussions, Tessa Baker, Baojiu Li, Antonio Maroto and Hans Winther for very useful discussions and comments on drafts of the paper and Alessandra Silvestri for very helpful correspondence and comments on a draft. JN and PGF were supported by Leverhulme, STFC, BIPAC and the Oxford Martin School.
\end{acknowledgments}



\appendix

\section{Comparing conventions for $f(R)$}
\label{CT}
Let us briefly review the mapping between $f(R)$ theories and chameleons, pointing out some important subtleties between different, typically-used conventions. In order to do so we establish a dictionary between the convention (largely) used in the literature for structure formation in $f(R)$ models (\eg{} \cite{0611321,0705.1158}) and that used in  chameleon phenomenology and screening effects (\eg{} \cite{1210.6880}).  The former convention we label I and the latter II: this paper uses convention I.  In order to avoid confusion when comparing with other literature, we here explicitly spell out these conventions and the mapping between them. 

{\bf Convention I:}  As we saw at the start of this section, the $f(R)$ action can be written (in the Jordan frame) as
\begin{equation}
S_J =\frac{1}{2}\int d^4 x\sqrt{-g}\, \left[R+f(R)\right] + \int d^4 x\sqrt{-g}\, {\cal L}_{\rm m}[\Phi_i,g_{\mu\nu}] \ ,
\end{equation}
which is then mapped into the equivalent Einstein frame scalar-tensor theory
\begin{eqnarray}
\nn S_E &=&\frac{1}{2}\int d^4 x\sqrt{-{\tilde g}}\, {\tilde R} \\
&+& \int d^4 x\sqrt{-{\tilde g}}\, 
\left[-\frac{1}{2}{\tilde g}^{\mu\nu}\tilde{\nabla}_{\mu}\phi \tilde{\nabla}_{\nu}\phi -V(\phi)\right] \nonumber
\\ &+& S_{matter}[\Phi_i, e^{-\beta \phi} {\tilde g}_{\mu\nu}]
\end{eqnarray}
where we have employed a conformal transformation
\begin{equation}
{\tilde g}_{\mu\nu} = e^{\beta \phi} g_{\mu\nu},
\end{equation}
and defined the field $\phi$ via
\begin{eqnarray}
1+f_{R} &=& e^{2 \phi \beta}.
\end{eqnarray} 
$\beta$ in this convention is $\sqrt{2/3}$.
The potential $V(\phi)$ is determined by
\begin{equation}
V(\phi)=\frac{1}{2}\frac{R f_{R} - f}{(1+f_{R})^{2}}.
\end{equation}

{\bf Convention II: }  The action we start with now is
\begin{equation}
S_J =\frac{1}{2}\int d^4 x\sqrt{-g}\, \left[f(R)\right] + \int d^4 x\sqrt{-g}\, {\cal L}_{\rm m}[\Phi_i,g_{\mu\nu}] \ ,
\end{equation}
i.e. $f(R)^{(II)} = R + f(R)^{(I)}$ where the Roman index denotes the convention. The metric ${\tilde g}_{\mu\nu}$ and the field $\phi$ are now defined via
\begin{eqnarray}
f_{R}^{(II)} &=& e^{- 2 \hat \beta \phi}, \\
{\tilde g}_{\mu\nu} &=& e^{- 2 \hat \beta \phi}  g_{\mu \nu} =  f_{R}^{(II)} g_{\mu \nu}.
\end{eqnarray}
$\hat \beta$ in this convention is $\sqrt{1/6}$.  Finally the potential $V(\phi)$ in the second convention is
\begin{equation}
V(\phi)^{(II)}=\frac{1}{2}\frac{R f_{R}^{(II)} - f^{(II)}}{(f_{R}^{(II)})^{2}}.
\end{equation}

{\bf The mapping: }
It is now clear that the difference between the two conventions boils down to a a field redefinition of $\phi$, namely
\begin{equation}
\phi_{(I)} \leftrightarrow - \phi_{(II)}.
\end{equation}
This means care has to be taken when considering which potentials have the correct properties to give rise to an effective chameleon.

\bibliographystyle{apsrev4-1}
\bibliography{QS-extended}
\end{document}